\documentclass[fleqn,usenatbib]{mnras}

\usepackage{newtxtext,newtxmath}

\usepackage[T1]{fontenc}
\usepackage{threeparttable}
\usepackage{ulem}

\DeclareRobustCommand{\VAN}[3]{#2}
\let\VANthebibliography\thebibliography
\def\thebibliography{\DeclareRobustCommand{\VAN}[3]{##3}\VANthebibliography}

\usepackage{graphicx}	
\usepackage{amsmath}	

\title[NGC 188 Red Giants]{WIYN Open Cluster Study: The Old Open Cluster, NGC 188, and a Re-evaluation of Lithium-Richness Among Red Giants}

\author[Q. Sun et al.]{
Qinghui Sun,$^{1,2,6}$
Constantine P. Deliyannis,$^{1,6}$
Bruce A. Twarog,$^{3}$
Barbara J. Anthony-Twarog,$^{3}$
\newauthor
Jeffrey D. Cummings,$^{4}$
and Aaron Steinhauer$^{5}$
\\
$^{1}$Department of Astronomy, Indiana University, Bloomington, IN 47405, USA\\
$^{2}$Currently at Department of Astronomy, Tsinghua University, Beijing, 100084, China\\
$^{3}$Department of Physics and Astronomy, University of Kansas, Lawrence, KS 660045, USA\\
$^{4}$Center for Astrophysical Sciences, Johns Hopkins University, Baltimore, MD 21218, USA\\
$^{5}$Department of Physics and Astronomy, State University of New York, Geneso, NY 14454, USA\\
$^{6}$Visiting Astronomer, Kitt Peak National Observatory, National Optical Astronomy Observatory, which is operated by the Association of Universities for \\ Research in Astronomy (AURA) under cooperative agreement with the National Science Foundation.\\
}

\date{Accepted 2022 April 28. Received 2022 April 7; in original form 2021 September 14}

\pubyear{2021}

\begin{document}
\label{firstpage}
\pagerange{\pageref{firstpage}--\pageref{lastpage}}
\maketitle

\begin{abstract}
We present WIYN\footnote//Hydra spectra of 34 red giant candidate members of NGC 188, which, together with WOCS\footnote \ \ and Gaia data yield 23 single members, 6 binary members, 4 single nonmembers, and 1 binary nonmember. We report [Fe/H] for 29 members and derive [Fe/H]$_{\rm{NGC188}}$ = +0.064 $\pm$ 0.018 dex ($\sigma_{\mu}$) (sky spectra yield A(Fe)$_{\odot}$ = 7.520 $\pm$ 0.015 dex ($\sigma_{\mu}$)). We discuss effects on the derived parameters of varying Yale-Yonsei isochrones to fit the turnoff. We take advantage of the coolest, lowest-gravity giants to refine the line list near Li 6707.8 \AA. Using synthesis we derive detections of A(Li)\footnote\ \ = 1.17, 1.65, 2.04, and 0.60 dex for stars 4346, 4705, 5027, and 6353, respectively, and 3$\sigma$ upper-limits for the other members. Whereas only two of the detections meet the traditional criterion for ``Li-richness'' of A(Li) $>$ 1.5 dex, we argue that since the cluster A(Li) vanish as subgiants evolve to the base of the RGB, all four stars are Li-rich in this cluster's context. An incidence of even a few Li-rich stars in a sample of 29 stars is far higher than what recent large surveys have found in the field. All four stars lie either slightly or substantially away from the cluster fiducial sequence, possibly providing clues about their Li-richness. We discuss a number of possibilities for the origin for the Li in each star, and suggest potentially discriminating future observations.

\end{abstract}


\footnotetext[1]{The WIYN Observatory is a joint facility of the University of Wisconsin-Madison, Indiana University, the National Optical Astronomy Observatory and the University of Missouri.}
\footnotetext[2]{WIYN Open Cluster Study, Mathieu\citet{Mathieu00}.}
\footnotetext[3]{A(Li) = 12 + log(N$_{Li}$/N$_H$), where N$_X$ is number of atoms of species X.}



\section{Introduction}

Lithium (Li) was first created during Big Bang Nucleosynthesis (Coc et al.\citealt{Coc14}, Boesgaard et al.\citealt{Boesgaard05a}), and is subsequently produced inside red giant stars, during nova and supernova explosions, and by spallation processes in the ISM or stellar surfaces (Travaglio et al.\citealt{Travaglio01}, Guiglion et al.\citealt{Guiglion19}). Through these (and potentially other) processes, the Milky Way has increased the Big Bang A(Li) from the predicted $\sim$ 2.7 to $\sim$ 3.3 dex observed in meteorites\citet[Anders \& Grevesse][]{Anders89}, young star clusters and the ISM\citet[Cummings et al.][= C17]{Cummings17}. Conversely, stars deplete Li throughout their lives. At minimum, observed surface A(Li) qualitatively follow the standard stellar theory (no rotation, diffusion, magnetic fields, mass loss, etc.; \citealt[Deliyannis][]{Deliyannis90a}; \citealt[Deliyannis et al.][= D90]{Deliyannis90b}) predictions that: a) nuclear reactions destroy Li at the base of the surface convection zone (SCZ) during pre-main sequence and, b) subgiant deepening of the SCZ results in dilution. Furthermore, considerable evidence suggests that rotational mixing depletes the surface A(Li) throughout the main sequence (MS) and possibly beyond for the full range of masses where Li is observed (Deliyannis et al.\citealt{Deliyannis19}, C17). By the time stars leave the MS, most are already severely Li-depleted and dilution (and perhaps rotational mixing) reduces the surface A(Li) further. First ascent red giants (``RGB'') and further evolved red giants should thus have little of their original Li preserved\citet[Anthony-Twarog et al.][]{Anthony21}. Consistent with these expectations, surveys show that the vast majority of red giants have no measurable Li, with detections or upper limits typically at levels (Brown et al.\citealt{Brown89}; Kumar et al.\citealt{Kumar11}; Casey et al.\citealt{Casey16}; Martell \& Shetrone\citealt{Martell13}) much lower than is observed on the MS.

However, a small fraction of giants have surprisingly high A(Li). Subgiant dilution is expected to reduce the surface A(Li) of Pop I stars by 1.8 -- 1.9 dex (see Fig. 14 in Charbonnel \& Lagarde\citealt{Charbonnel10}), depending slightly on mass, and of Pop II stars by $\sim 1.2$ dex (D90).  If solar-metallicity stars form with A(Li) $\sim$ 3.3 dex, then after subgiant dilution (``first dredge-up''), Pop I giants should have at most A(Li) = 1.5 dex, and likely less if rotational mixing and other mechanisms have depleted the internal Li abundance during MS evolution. Red giants with A(Li) $>$ 1.5 dex are thus assumed to have enriched their surfaces with Li, and have often been labelled as ``Li-rich'', although the boundary of 1.5 dex may depend on metallicity and other circumstances. Recent large surveys have identified 0.64\% of Pop I red giants below $2M_{\odot}$\citet[Deepak \& Reddy][from the GALAH survey]{Deepak19} and 1.29\% of all giants\citet[Gao et al.][]{Gao19} as Li-rich, while Kirby et al.\citet{Kirby16} tag 0.3$\pm$0.1\% of globular cluster giants as such.  

Among proposed mechanisms that might enhance the surface Li of giants, a prominent one is the ``$^7$Be-transport mechanism''\citet[Cameron \& Fowler][]{Cameron71}, that may apply where the nucleosynthetic regions mix briefly with the surface. The situations may include the He-core flash at the tip of the RGB, thermohaline mixing as stars approach the tip of the RGB\citet[Lattanzio et al.][]{Lattanzio15}, evolution past the Thomas Peak (Thomas\citealt{Thomas67}, Demarque\citealt{Demarque87}), also known as the RGB bump or luminosity bump on the RGB (Charbonnel \& Balachandran\citealt{Charbonnel00}) and its variant, the Denissenkov\citet{Denissenkov12} mechanism, for stars that were very rapid rotators on the MS, and merger of a helium white dwarf (HeWD) with another HeWD or RGB star\citet[Zhang et al.][]{Zhang20}. In addition to the $^7$Be-transport mechanism, purely external Li enrichment mechanisms have been proposed, and may be more likely to be responsible for Li-rich giants that have not yet evolved to the luminosity bump. Examples include Li enrichment by engulfment of a sub-stellar companion (e.g. a planet or brown dwarf \citealt[Alexander ][]{Alexander67}; \citealt[Ashwell et al.][]{Ashwell05}; \citealt[Siess \& Livio ][]{Siess99}), mass transfer from a Li-rich asymptotic giant branch (AGB) companion \citet[Aguilera-Gomez et al.][]{Aguilera16a} or nova companion\citet[Jose \& Hernanz][]{Jose98}.

Although the number of giants that can be studied in open clusters is relatively small, compared to surveys like GALAH, open clusters offer some important advantages. In particular, their ages can be determined more reliably than for field stars, providing potentially critical parameters such as the RGB stellar mass. Since stars in an open cluster all have the same age and initial composition, Li-rich stars in an open cluster can be compared to otherwise similar stars, supplying insight into the origin of the anomalous behavior. For example, Anthony-Twarog et al.\citet{Anthony13} discovered a Li-rich (A(Li) = 2.3 dex) red giant member (WOCS 7017) of the open cluster NGC 6819. The star is located below the red clump and redward of the RGB in the color-magnitude diagram (CMD), has asteroseismically-inferred anomalously lower mass \citet[Handberg et al.][]{Handberg17}, and is consistent with the $^7$Be-transport mechanism having created this star's Li in a He core flash.

To gain further insight about Li enrichment of red giants, we study Li in the giants of one of the older open clusters in the Milky Way -- NGC 188. One critical factor of interest is that this cluster's turnoff stars are late F dwarfs on the cool side of the Li Dip, so their surface Li is already significantly depleted before they leave the MS. The Li Dip is a severe depletion of Li in F dwarfs, first discovered in the Hyades\citet[Boesgaard \& Tripicco][]{Boesgaard86}, shown to occur during the MS\citet[Boesgaard et al.][]{Boesgaard88}, and most likely due to rotational mixing induced by angular momentum loss (C17; see Section \ref{sec:explanation}). Subgiants of NGC 188 show sharply declining Li as they evolve to lower $T_{eff}$, reaching undetectable levels at the base of the RGB\citet[Deliyannis et al.][in preparation]{Deliyannis22}. This suggests that any red giant member observed to have A(Li) higher than the upper limits observed at the base of the RGB is enriched either internally or externally, or is an abnormally good preserver of Li.

This paper is arranged as follows. Section \ref{sec:obs} discusses selection of targets and the observations; Section \ref{sec:rv} presents radial-velocity measurements and evaluation of binarity and membership; Section \ref{sec:iron_abun} describes determination of iron abundances of each star and the cluster; Section \ref{sec:li_abun} describes determination of Li abundances; Section \ref{sec:explanation} discusses possible interpretations of the Li abundances; and Section \ref{sec:summary} provides the paper summary.

\section{OBSERVATIONS AND DATA REDUCTIONS} \label{sec:obs}

Data were taken using the Hydra MOS on the WIYN 3.5m telescope in a single configuration of the red cable during the nights of 20 and 21 February 2017 (hereafter n1 and n2), centered on R.A. = 0$^h$47$^m$23.962$^s$, Dec = 85$^\circ$13$\arcmin$45.94$\arcsec$. The 316@63.4 echelle grating was used in order 8 with the X19 filter, providing a wavelength coverage from 6428 to 6847 \AA, a dispersion of 0.205 \AA\ pixel$^{-1}$, and R $\sim$ 17000 as measured from the arcs. Targets were selected from the proper-motion study of Platais et al.\citet[][=P03]{Platais03}, which covers a 0.75 $deg^2$ area. All 29 stars on or near the red giants branches with $V < 14.12$ mag and proper motion probability ($P_{\mu}$) greater than 70\% were placed on fibers (20 stars have $P_{\mu} \geq 94\%$). Since there were fibers to spare, another 6 stars with $10\% \leq P_{\mu} \leq 70\%$ were placed on fibers; 4 of these are on or near the red giant branches and 2 are bluer. One star (ID=4150 from P03) contributed negligible flux and is omitted from further consideration. Table \ref{tab:all star} shows the ID, R.A., Dec, $V$, $B-V$, and $P_{\mu}$, taken from P03, of the remaining 34 program cluster stars. These stars are shown on a CMD in Section \ref{sec:iron_abun}. 

Biases, dome flats, Th-Ar arcs, and daytime spectra of clear blue sky were taken in the same cluster configuration. Fibers not assigned to stars were positioned to measure (night) sky background. The radial-velocity standard HD 144579 was observed on night 1 to provide an external check on the wavelength calibration. Exposures were taken of the NGC 188 configuration of 45 and 37.5 min during n1, and of 40 and 35 min on n2. We followed the same processing and data reduction procedures as in Sun et al.\citet{Sun20}.

\begin{table*}
	\centering
	\caption{Parameters for NGC 188 RGB stars}
	\label{tab:all star}
	\begin{threeparttable}
	\resizebox{2.\columnwidth}{!}{%
	\begin{tabular}{ccccccccccccccccccccc}
		\hline
			Id$^1$ & R.A.$^1$ & dec$^1$ & V$^2$ & B-V$^2$ & $V_{RAD}^3$ & $\sigma^3$ & $V_{RAD}^4$ & $\sigma/i^4$ & {\it v} sin {\it i}$^5$ & $\sigma^5$ & $P_{V_{RAD}}^4$ & $P_{\mu}^1$ & mm$^4$ & $\pi^6$ & $\sigma^6$ & $\mu_{\alpha}^6$ & $\sigma^6$ & $\mu_{\delta}^6$ & $\sigma^6$ & mm$^7$ \\			  
			&  &  &  &  & ours &  & G08 &  & ours &  & G08 & P03 & G08 & DR2 &  & DR2 &  & DR2 &  & ours \\
			& $^h\ ^m\ ^s$ & $^\circ\ \arcmin\ \arcsec$ & mag & mag & km s$^{-1}$ & & km s$^{-1}$ &  & km s$^{-1}$ & & \% & \% &  & mas & & mas yr$^{-1}$ & & mas yr$^{-1}$ & & final \\ 

		\hline
		1141 & 0 44 44.499 & 85 32 16.42 & 14.131 & 1.181 & -43.53 & 0.94 & -42.55 & 0.59 & $<$ 15 & -- & 98 & 88 & SM & 0.5287 & 0.0162 & -2.383 & 0.026 & -0.927 & 0.024 & SM \\
		3015 & 0 32 00.732 & 85 11 46.40 & 12.649 & 1.271 & -43.19 & 0.97 & -42.62 & 1.05 & $<$ 15 & -- & 98 & 24 & SM & 0.5200 & 0.0195 & -2.344 & 0.037 & -1.108 & 0.039 & SM  \\
		3062 & 0 34 11.416 & 85 26 10.92 & 12.983 & 1.095 & -42.88 & 0.52 & -42.34 & 0.40 & $<$ 15 & -- & 98 & 20 & SM & 0.3630 & 0.0198 & -2.136 & 0.037 & -2.844 & 0.039 & SN \\
		3271 & 0 35 09.141 & 85 17 16.91 & 12.437 & 1.141 & -42.24 & 0.72 & -42.28 & 0.54 & $<$ 15 & -- & 98 & 94 & SM & 0.4853 & 0.0250 & -2.278 & 0.045 & -1.204 & 0.045 & SM  \\
		4228 & 0 42 22.988 & 85 20 33.70 & 13.898 & 1.093 & -43.77 & 0.57 & -43.56 & 0.65 & $<$ 15 & -- & 95 & 98 & SM & 0.4984 & 0.0153 & -2.370 & 0.030 & -1.122 & 0.023 & SM \\
		4294 & 0 42 25.541 & 85 16 22.03 & 12.936 & 1.192 & -40.45 & 1.09 & -42.57 & 1.39 & $<$ 15 & -- & 98 & 96 & SM & 0.5549 & 0.0264 & -2.031 & 0.046 & -0.819 & 0.037 & SM  \\
		4346 & 0 42 35.264 & 85 13 25.63 & 13.642 & 0.963 & -41.50 & 0.65 & -41.35 & 0.96 & $<$ 15 & -- & 95 & 98 & SM & 0.4793 & 0.0153 & -2.303 & 0.028 & -0.938 & 0.023 & SM \\
		4408 & 0 43 26.792 & 85 09 17.74 & 13.148 & 1.083 & -54.09 & 0.55 & -53.40 & 1.25 & $<$ 15 & -- &  0 & 34 & SN & 0.9499 & 0.0232 & -1.747 & 0.042 & -0.203 & 0.034 & SN  \\
		4524 & 0 46 19.656 & 85 20 08.67 & 12.445 & 1.180 & -38.53 & 1.75 & -42.97 & 6.61 & $<$ 15 & -- & 98 & 97 & BM & 0.4882 & 0.0242 & -2.138 & 0.051 & -1.234 & 0.037 & BM  \\
		4565 & 0 45 51.060 & 85 18 08.28 & 12.408 & 1.304 & -43.87 & 1.26 & -42.18 & 7.33 & $<$ 15 & -- & 98 & 96 & BM & 0.6207 & 0.0251 & -2.388 & 0.049 & -1.032 & 0.043 & BM  \\
		4668 & 0 44 52.370 & 85 14 05.66 & 12.314 & 1.295 & -41.45 & 1.11 & -41.44 & 1.07 & $<$ 15 & -- & 96 & 98 & SM & 0.5058 & 0.0246 & -2.235 & 0.045 & -0.946 & 0.043 & SM  \\
		4705 & 0 45 22.477 & 85 12 38.27 & 13.894 & 0.944 & -40.10 & 0.77 & -42.53 & 99.99 & 17.7 & 0.5 & 98 & 98 & BM & 0.5018 & 0.0165 & -2.296 & 0.034 & -0.961 & 0.027 & BM \\
		4756 & 0 44 12.250 & 85 09 31.33 & 11.343 & 1.536 & -42.04 & 1.17 & -42.40 & 0.69 & $<$ 15 & -- & 98 & 71 & SM & 0.5219 & 0.0453 & -2.389 & 0.090 & -0.965 & 0.061 & SM  \\
		4829 & 0 47 29.668 & 85 24 14.09 & 12.797 & 1.173 & -41.69 & 1.05 & -41.64 & 1.01 & $<$ 15 & -- & 97 & 98 & SM & 0.4876 & 0.0252 & -2.372 & 0.045 & -0.947 & 0.038 & SM \\
		4843 & 0 46 39.207 & 85 23 33.74 & 11.512 & 1.341 & -48.20 & 1.10 & -42.08 & 6.97 & $<$ 15 & -- & 98 & 75 & BM & 0.5309 & 0.0236 & -2.592 & 0.044 & -0.700 & 0.038 & BM \\
		4909 & 0 47 18.356 & 85 19 45.72 & 12.975 & 1.211 & -43.55 & 1.04 & -42.95 & 1.01 & $<$ 15 & -- & 98 & 96 & SM & 0.4587 & 0.0234 & -2.426 & 0.046 & -0.932 & 0.035 & SM \\
		5027 & 0 47 56.955 & 85 14 55.98 & 11.938 & 1.097 & -42.73 & 1.75 & -42.23 & 1.51 & 22.1 & 1.1 & 98 & 96 & SM & 0.4512 & 0.0246 & -2.322 & 0.050 & -0.940 & 0.036 & SM \\
		5048 & 0 46 33.623 & 85 14 34.32 & 13.858 & 1.105 & -43.19 & 0.98 & -42.27 & 15.86 & $<$ 15 & -- & 98 & 94 & BM & 0.4812 & 0.0176 & -2.212 & 0.031 & -0.803 & 0.025 & BM \\
		5085 & 0 46 59.570 & 85 13 15.77 & 12.347 & 1.195 & -42.15 & 1.12 & -42.09 & 0.63 & $<$ 15 & -- & 98 & 98 & SM & 0.4918 & 0.0275 & -2.275 & 0.051 & -0.939 & 0.038 & SM  \\
		5133 & 0 47 32.099 & 85 11 02.52 & 14.077 & 1.085 & -43.55 & 0.59 & -43.40 & 1.45 & $<$ 15 & -- & 96 & 98 & SM & 0.5052 & 0.0178 & -2.320 & 0.035 & -0.976 & 0.026 & SM \\
		5438 & 0 48 58.958 & 85 12 29.58 & 13.636 & 1.115 & -43.25 & 0.89 & -41.30 & 0.99 & $<$ 15 & -- & 95 & 98 & SM & 0.5095 & 0.0142 & -2.133 & 0.030 & -0.988 & 0.022 & SM \\
		5597 & 0 53 35.537 & 85 20 58.29 & 12.875 & 1.237 & -43.30 & 1.03 & -43.04 & 1.14 & $<$ 15 & -- & 98 & 26 & SM & 0.4771 & 0.0215 & -2.267 & 0.041 & -0.969 & 0.036 & SM  \\
		5835 & 0 49 05.522 & 85 26 07.74 & 12.715 & 1.265 & -43.25 & 1.03 & -42.66 & 0.76 & $<$ 15 & -- & 98 & 94 & SM & 0.5077 & 0.0219 & -2.441 & 0.044 & -0.894 & 0.035 & SM  \\
		5855 & 0 51 21.624 & 85 12 37.71 & 13.370 & 1.162 & -44.68 & 0.97 & -42.35 & 3.74 & $<$ 15 & -- & 98 & 98 & BM & 0.4730 & 0.0147 & -2.433 & 0.032 & -0.864 & 0.023 & BM  \\
		5894 & 0 46 29.900 & 85 11 51.98 & 14.054 & 1.100 & -43.90 & 0.67 & -43.19 & 0.59 & $<$ 15 & -- & 97 & 96 & SM & 0.4958 & 0.0171 & -2.340 & 0.034 & -1.144 & 0.025 & SM \\
		6353 & 0 57 18.355 & 85 10 28.77 & 12.498 & 1.227 & -42.72 & 1.03 & -42.11 & 1.01 & $<$ 15 & -- & 98 & 89 & SM & 0.5015 & 0.0207 & -2.338 & 0.046 & -0.866 & 0.035 & SM  \\
		6188 & 0 53 34.899 & 85 11 14.53 & 13.395 & 1.168 & -43.21 & 1.15 & -42.67 & 0.63 & $<$ 15 & -- & 98 & 98 & SM & 0.4953 & 0.0128 & -2.324 & 0.027 & -0.893 & 0.021 & SM  \\
		6602 & 1 02 52.500 & 85 17 56.06 & 12.689 & 1.357 & -43.08 & 0.99 & -42.33 & 0.87 & $<$ 15 & -- & 98 & 35 & SM & 0.5015 & 0.0188 & -2.404 & 0.045 & -0.783 & 0.032 & SM  \\
		6619 & 1 01 35.198 & 85 15 09.87 & 13.745 & 1.301 & -80.22 & 0.69 & -80.50 & 0.91 & $<$ 15 & -- &  0 & 81 & SN & 0.2482 & 0.0145 & -2.949 & 0.029 & -0.093 & 0.023 & SN \\
		6719 & 1 05 04.797 & 85 22 23.34 & 12.998 & 0.646 & -29.58 & 0.81 & -32.98 & 14.94 & $<$ 15 & -- &  0 & 10 & BU & 0.8859 & 0.0253 & -2.806 & 0.054 & -2.910 & 0.043 & BN \\
		6982 & 0 54 11.362 & 85 15 23.14 & 12.409 & 1.207 & -42.58 & 1.15 & -42.14 & 0.74 & $<$ 15 & -- & 98 & 74 & SM & 0.5193 & 0.0224 & -2.329 & 0.051 & -0.872 & 0.038 & SM \\
		8129 & 0 41 19.728 & 85 01 27.00 & 13.469 & 1.135 & -42.91 & 0.67 & -42.88 & 1.25 & $<$ 15 & -- & 98 & 95 & SM & 0.4993 & 0.0145 & -2.161 & 0.027 & -1.081 & 0.023 & SM \\
		9159 & 0 54 36.497 & 85 01 15.32 & 12.472 & 1.174 & -42.86 & 1.06 & -42.83 & 0.55 & $<$ 15 & -- & 98 & 88 & SM & 0.4777 & 0.0226 & -2.297 & 0.049 & -0.905 & 0.038 & SM \\
		9291$^8$ & 0 56 32.806 & 84 58 40.65 & 12.110 & 0.972 & 2.93 & 0.56 & -- & -- &  $<$ 15 & -- & -- & 53 & -- & 0.5530 & 0.0208 & -1.703 & 0.042 & -1.347 & 0.034 & SN \\
		\hline
		\end{tabular}
	}
	\begin{tablenotes}
		\item[1] ID, R.A., dec, and proper motions membership probability $P_{\mu}$ from P03.
		\item[2] $V$ band magnitude and $B-V$ color from H00. If H00 did not observe the star, we use the average $V$ and $B-V$ instead (see text).
		\item[3] Our measurement of radial velocity ($V_{RAD}$) and errors from {\it fxcor} using the combined spectra.
		\item[4] $V_{RAD}$ and $\sigma$ from Geller et al.\citet{Geller08} (hereafter G08), where errors are standard deviation to precision ratio. $P_{V_{RAD}}$ are membership probabilities \\based on $V_{RAD}$ from G08. mm are final determination of multiplicity and membership from G08. The first letter ``S'' means single, ``B'' means binary. \\The second letter ``M'' means a member, ``N'' means nonmember, and ``U'' means G08 was uncertain about the star's membership.
		\item[5] Rotational velocity ({\it v} sin {\it i}) and errors from {\it fxcor}. For stars below our resolution limit, we substitute the {\it v} sin {\it i} reported by {\it fxcor} with an upper limit of 15 km $s^{-1}$.
		\item[6] Gaia parallax, proper motion in R.A. and Dec, and their standard errors.
		\item[7] Our final multiplicity and membership.
		\item[8] G08 has not observed star 9291.
	\end{tablenotes}
		\end{threeparttable}
\end{table*}

\section{Radial Velocities and Membership}  \label{sec:rv}

\subsection{Radial Velocity}

\begin{figure}
	\centering
	\includegraphics[width=0.5\textwidth]{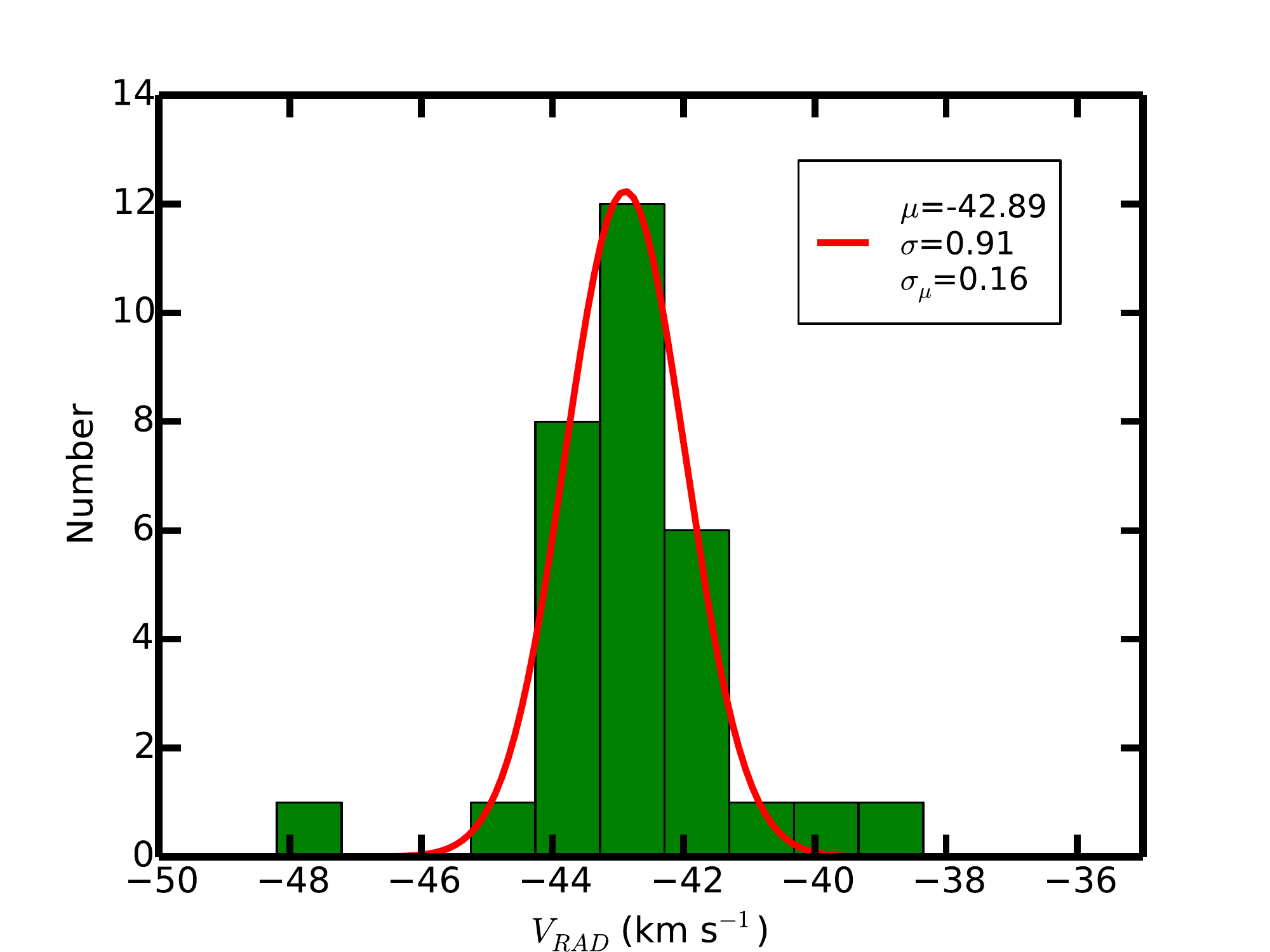}
	\caption{Radial-velocity distribution for red giant stars in NGC 188.}
	\label{fig:rad}
\end{figure}

We used the IRAF\footnote{IRAF is distributed by the National Optical Astronomy Observatories, which are operated by the Association of Universities for Research in Astronomy Inc., under cooperative agreement with the National Science Foundation.} task {\it fxcor} to measure radial velocities ($V_{RAD}$) for all the spectral lines of each star, fit a Gaussian profile to the $V_{RAD}$ distribution in Fourier space, and then adopted the mean value of the Gaussian profile as the star's $V_{RAD}$. Using methods described in C17 (for additional details see Cummings \citealt{Cummings11} and Steinhauer\citealt{Steinhauer03}), comparisons were made to template spectra of radial velocity standards that have {\it v} sin {\it i} $<$ 5 km s$^{-1}$, that were observed with the same instrument and similar resolution (with typical SNR in the hundreds), and that span a wide range of spectral types. We derived $V_{RAD}$ = -59.64 $\pm$ 0.72 km s$^{-1}$ for the radial-velocity standard HD 144579 (observed during n1 only), in superb agreement with the value -59.38 $\pm$ 0.015 km s$^{-1}$ from Soubiran et al.\citet{Soubiran13}. The $V_{RAD}$ for individual stars were measured independently for each night. We found a 0.84 km s$^{-1}$ systematic shift of n2 relative to n1. We thus shifted the spectra of n2 by -0.018\ \AA\ to match n1 and co-added the shifted spectra from n2 to n1 to get final spectra with a higher signal-to-noise ratio (SNR), on which we ran the {\it fxcor} task once again. Table \ref{tab:all star} shows our $V_{RAD}$ measurements and errors from the final co-added spectra. Figure \ref{fig:rad} shows the $V_{RAD}$ distribution; typical errors for individual stars are 0.5--1.1 km s$^{-1}$. A Gaussian fit yields a mean $V_{RAD}$ = -42.89 $\pm$ 0.16 km s$^{-1}$ ($\sigma_{\mu}$, $\sigma$ = 0.93 km s$^{-1}$), and we initially classified stars within 2 $\sigma$ of the mean $V_{RAD}$ (in the range [-44.75, -41.03] km s$^{-1}$) as $V_{RAD}$ probable members. Table \ref{tab:all star} also shows our derived rotational velocities ({\it v} sin {\it i}) and errors calculated from {\it fxcor} based on line broadening. Our resolution suggests treating {\it v} sin {\it i} below $\sim$ 15 km s$^{-1}$ as upper limits, so all {\it fxcor}-derived values less than 15 km s$^{-1}$ are reported in Table \ref{tab:all star} as ``$<$ 15''.

\begin{figure}
	\centering
	\includegraphics[width=0.5\textwidth]{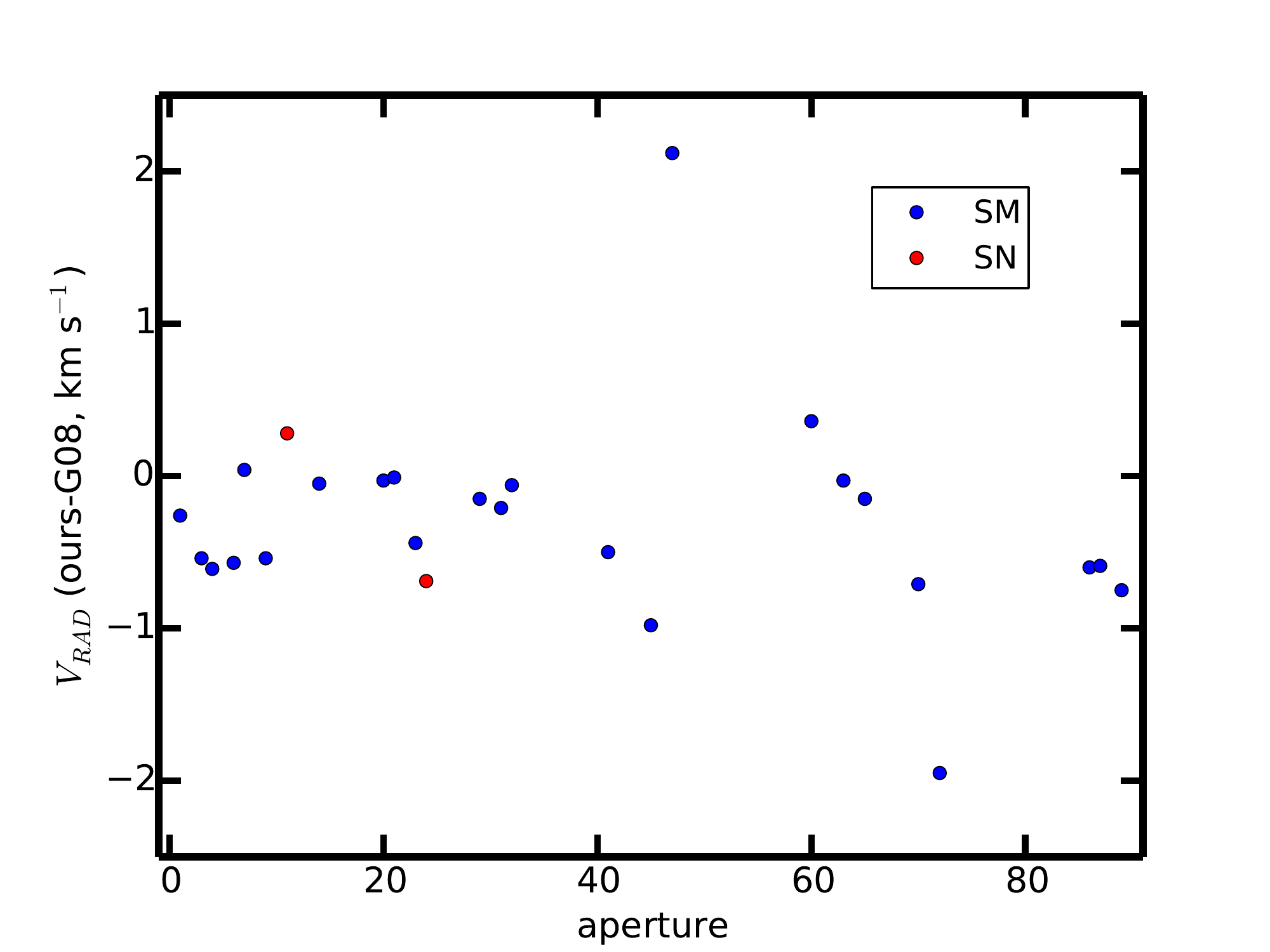}
	\caption{Comparison of our $V_{RAD}$ to those of G08 for stars designated as single by G08. The Single Members (SM) are shown as blue dots, and the two Single Nonmembers (SN) as red dots.}
	\label{fig:rad_single}
\end{figure}

In their thorough radial-velocity study of NGC 188, Geller et al.\citet[G08]{Geller08} took repeated measurements of 1046 stars in the direction of NGC 188, and found 473 probable cluster members. Their average cluster $V_{RAD}$ calculated from single stars in common with our study is $-42.52 \pm 0.12$ km s$^{-1}$ ($\sigma_{\mu}$). This is within 2.3 $\sigma_{\mu}$ (our $\sigma_{\mu}$) of our value. All of our stars except star 9291 were observed by G08, and the $V_{RAD}$ and membership information from G08 are included in Table \ref{tab:all star}.  Figure \ref{fig:rad_single} shows the difference between our values of $V_{RAD}$ and those of G08 for stars identified as single by G08 (SM = Single Member, SN = Single Nonmember). The mean difference is $\mu_d$ = -0.293 $\pm$ 0.135 km s$^{-1}$ ($\sigma_{\mu}$). After excluding stars 4294 and 5438, which are outliers that fall beyond $2 \sigma$, the mean difference is $\mu_d$ = -0.324 $\pm$ 0.073 km s$^{-1}$ ($\sigma_{\mu}$). The differences for binary stars are larger, presumably because G08 report the mean binary $V_{RAD}$ derived from numerous individual measurements, whereas ours are instantaneous measurements.

\subsection{Membership}

\begin{figure*}
	\centering
	\includegraphics[width=1.0\textwidth]{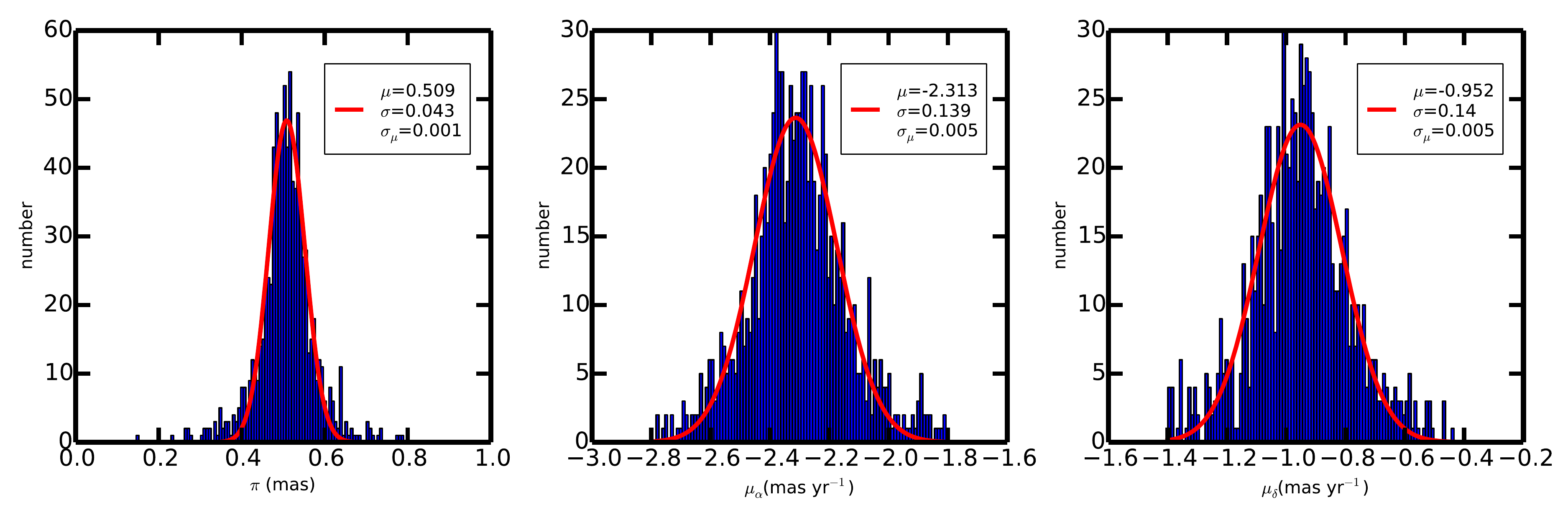}
	\caption{$\pi$ (mas), $\mu_{\alpha}$ (mas yr$^{-1}$), and $\mu_{\delta}$ (mas yr$^{-1}$) distribution. A Gaussian function is fitted (red solid line) to each histogram, and the mean ($\mu$), standard deviation ($\sigma$), standard deviation of the mean ($\sigma_{\mu}$) are annotated on the top right of each panel.}
	\label{fig:gaia_hist}
\end{figure*}

\begin{figure}
	\centering
	\includegraphics[width=0.5\textwidth]{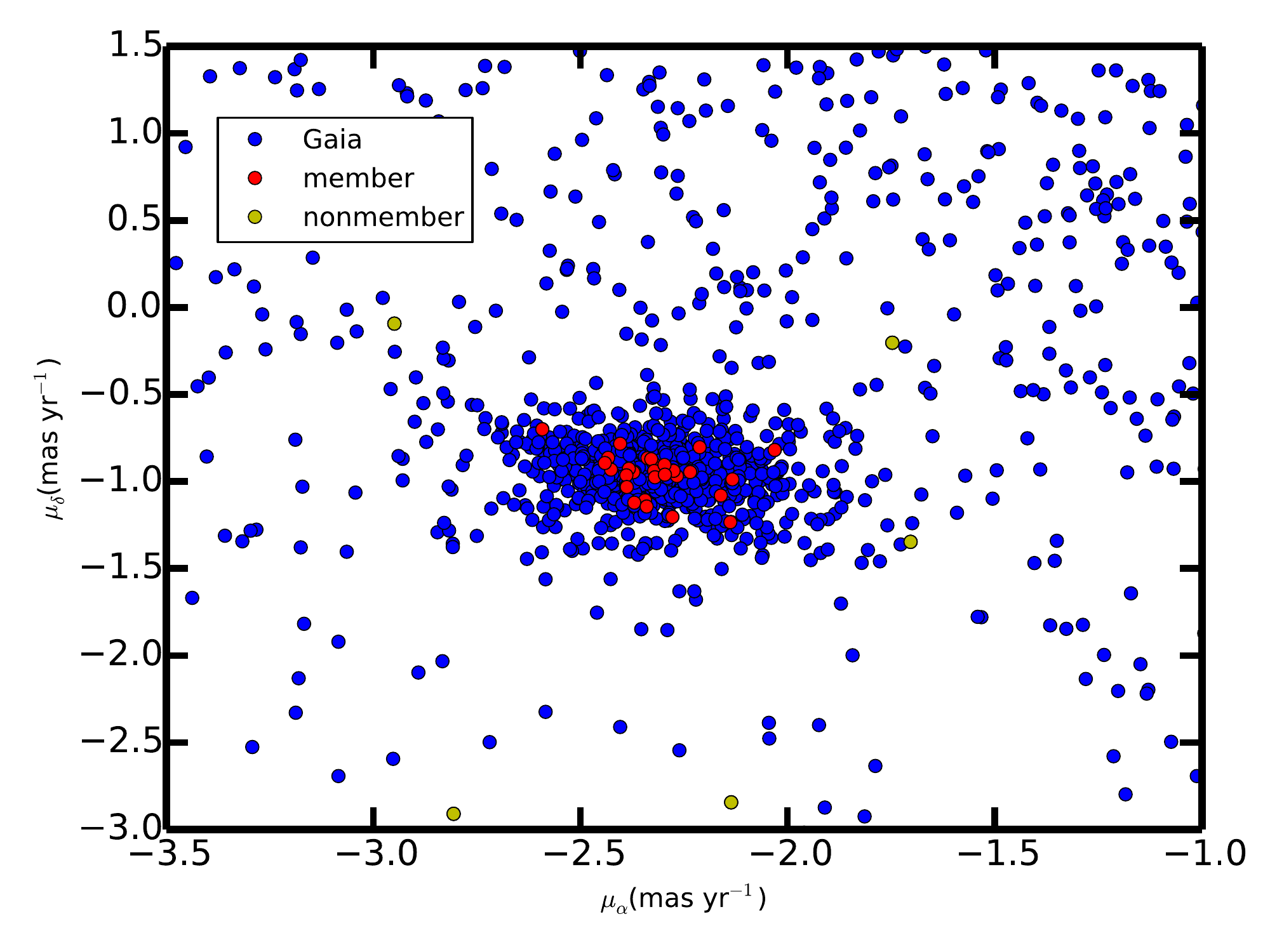}
	\caption{$\mu$ of our program stars from Gaia. Gaia stars are shown in blue disks, members are shown in red disks, and the five nonmembers are shown in yellow disks.}
	\label{fig:gaia_PV}
\end{figure}

As further checks of membership, especially for stars with $P_{\mu}$ from P03 in the range $10\% \leq P_{\mu} \leq 70\%$, we consider the $V_{RAD}$ in Table \ref{tab:all star} and information from Gaia DR2\citet[Gaia Collaboration][]{Gaia16, Gaia18} about parallax ($\pi$), $\mu$ in R.A. ($\mu_{\alpha}$) and $\mu$ in dec ($\mu_{\delta}$), shown in the last few columns of Table \ref{tab:all star}. The stars from Gaia lie within 0.55$^\circ$ of the center of our Hydra configuration and satisfy: $G\ <$ 18 mag, -2.8 mas yr$^{-1}$ $<\ \mu_{\alpha}\ <$ -1.8 mas yr$^{-1}$, -1.4 mas yr$^{-1}$ $<\ \mu_{\delta}\ <$ -0.4 mas yr$^{-1}$. Figure \ref{fig:gaia_hist} shows histograms for $\pi$, $\mu_{\alpha}$, and $\mu_{\delta}$ with Gaussians fit to the obvious cluster.  The four stars, 3062, 4408, 6619, and 6719, fall well outside the 2$\sigma$ ranges in $\pi$, $\mu_{\alpha}$, and $\mu_{\delta}$, as illustrated especially clearly in the point vector diagram of Figure \ref{fig:gaia_PV} for the last two of these criteria (yellow disks), so we label these stars as nonmembers. Star 4565 lies between 2$\sigma$ and 3$\sigma$ away from the mean in $\pi$ but is a clear member based on all other criteria so we label it as a member. Although star 9291 has $\pi$ consistent with membership, its $\mu_{\delta}$ lies between 2$\sigma$ and 3$\sigma$ away from the mean, its $\mu_{\alpha}$ is more than 3$\sigma$ away from the mean, and its (our) $V_{RAD}$ is inconsistent with single star membership, so we label it as a nonmember (5$^{th}$ yellow disk in Figure \ref{fig:gaia_PV}). We thus label 29 out of 34 stars as members. Regarding membership, we note the following points. Three of our final members had $P_{\mu}$ from P03 in the range $10\% \leq P_{\mu} \leq 70\%$, so including them in our configuration has enhanced our sample. Of the five nonmembers, four have $P_{\mu}$ from P03 in the range $10\% \leq P_{\mu} \leq 70\%$ but one has $P_{\mu}=81\%$. Three of the five nonmembers have $P_{V_{RAD}} = 0$ (from G08), but one has $P_{V_{RAD}} = 98\%$ (the 5$^{th}$ star is 9291, not measured by G08). This underscores the desirability of using multiple criteria for delineating membership whenever possible. Figure \ref{fig:Isochrone} shows all program stars on the CMD, with members as red disks and nonmembers as green disks. The membership designations from Cantat-Gaudin et al.\citet{Cantat18} based on the Gaia DR2 agree with all of ours except for star 9291, which they list as a member. Our spectrum shows no detectable Li in this star.

Regarding multiplicity, our power spectra from {\it fxcor} confirm single-star status for all stars labelled as such by G08, and indicate single-star status for star 9291 (not observed by G08). G08 designate seven of our Hydra stars as binaries, based on radial-velocity variations and orbital solutions. Of these, stars 4524, 4565, 4843, 5048, 5855 (all G08 Binary Members = BM) show double peaks from {\it fxcor} analysis of our spectra, confirming binarity. Star 6719, which we classify as a nonmember, does not show more than one peak in the power spectra; perhaps the secondary is faint. Star 4705, a definite binary from G08, also does not show a double/multiple peak; this very interesting star is discussed in detail in Section \ref{sec:explanation}. The only differences between our multiplicity/membership designations and those of G08 are for stars 3062 and 6719 (we designate them as nonmembers); star 9291 was not observed by G08. The last column in Table \ref{tab:all star} shows our final multiplicity/membership assignments. Figure \ref{fig:Isochrone} shows binaries as downward triangles and nonmembers as green disks. For the rest of the paper, we limit our discussion to the 29 cluster members.

\section{Stellar Parameters and Iron Abundances} \label{sec:iron_abun}

\subsection{B-V, Effective Temperature, Surface Gravity, and Microturbulence}     \label{sec:atmosphere}

We use the CCD photometry of Hainline et al.\citet[][H00, $BVRI$]{Hainline00}, of Sarajedini et al.\citet[S99]{Sarajedini99} ($UBVRI$), and of P03 ($BV$) to derive $T_{\rm{eff}}$. H00 covers a 40' $\times$ 40' region of NGC 188 that contains most of our member stars (26 out of 29), and S99 covers a 23' $\times$ 23' region that contains 20 out of the 29 members. Since H00 and S99 are on the same photometric scale for all filters to within the errors, we averaged the H00 and S99 $BVRI$ photometry for stars in common, used the H00 or S99 photometry for stars appearing only in one study, and used the S99 U photometry. The P03 photometry has a small offset and slope relative to H00 and S99, so for two stars (4829 and 1141) not observed by either H00 or S99, we adopted the P03 photometry shifted onto the H00 and S99 scale.  

To reduce statistical errors and to better sample the spectral energy distributions of our stars, we use all 10 possible color combinations from $UBVRI$ to derive an averaged effective $B-V$ for each star in NGC 188, as follows (advantages of this method are discussed in C17). Using only our stars, we plot each color against $B-V$, fit a polynomial, and then convert that color for each star into an effective $B-V$. We then average the nine transformed $B-V$ and the actual $B-V$ into an overall effective $B-V$ (hereafter simply ``$B-V$''). Some stars lack measurements in certain bands, so we use as many existing colors as possible. Table \ref{tab:atmosphere} shows the (average) $V$ magnitude, $B-V$, and $\sigma\ (B-V)$ (standard deviation of the 10 $B-V$ colors).

\begin{table*}
	\caption{Stellar atmosphere, metallicity, and Li abundance for 29 RGB member stars}
	\label{tab:atmosphere}
	\begin{threeparttable}
	\begin{tabular}{ccccccccccccc}
		\hline
			ID & $V_{avg}^1$ & $(B-V)_{avg}^2$ & $\sigma\ (B-V)$ $^2$ & $T_{\rm{eff}}^3$ & $\sigma(T_{eff})^3$ & log $g$ & $\xi$ & SNR$^4$ & [Fe/H]$^5$ & $\sigma^5$ & $\sigma_{\mu}^5$ & A(Li)$^6$ \\
			& mag & mag & mag & K & K & & km s$^{-1}$ & pixel$^{-1}$ & dex & dex & dex & dex \\
		
		\hline
		1141 & 14.131 & 1.181 & -- & 4655 & -- & 3.23 & 1.08 & 204 & -0.098 & 0.081 & 0.047 & $<$ 0.43 \\
		3015 & 12.649 & 1.271 & 0.015 & 4496 & 27 & 2.46 & 1.18 & 161 & 0.006 & 0.124 & 0.062 & $<$ 0.07 \\ 
		3271 & 12.453 & 1.130 & -- & 4751 & -- & 2.34 & 1.20 & 274 & 0.135 & 0.246 & 0.123 & $<$ 0.15 \\
		4228 & 13.921 & 1.048 & 0.016 & 4916 & 36 & 3.13 & 1.09 & 121 & 0.065 & 0.192 & 0.111 & $<$ 0.65 \\
		4294 & 12.952 & 1.150 & 0.025 & 4713 & 50 & 2.63 & 1.16 & 133 & 0.023 & 0.156 & 0.090 & $<$ 0.17\\
		4346 & 13.642 & 0.975 & 0.019 & 5076 & 46 & 3.00 & 1.11 & 153 & 0.044 & 0.218 & 0.109 & 1.17 $\pm$ 0.04  \\
		4524 & 12.445 & 1.180 & 0.028 & 4657 & 54 & 2.33 & 1.20 & 131 & 0.084 & 0.174 & 0.087 & $<$ -0.15 \\
		4565 & 12.430 & 1.293 & 0.013 & 4458 & 23 & 2.32 & 1.20 & 272 & 0.013 & 0.221 & 0.110 & $<$ -0.20 \\
		4668 & 12.314 & 1.295 & 0.012 & 4455 & 22 & 2.25 & 1.21 & 80 & 0.066 & 0.171 & 0.099 & $<$ 0.25 \\
		4705 & 13.918 & 0.967 & 0.039 & 5094 & 93 & 3.13 & 1.09 & 92 & 0.156 & 0.247 & 0.175 & 1.65 $\pm$ 0.06 \\
		4756 & 11.359 & 1.551 & 0.044 & 4039 & 64 & 1.53 & 1.30 & 543 & -0.012 & 0.149 & 0.075 & $<$ -1.20 \\
		4829 & 12.797 & 1.173 & -- & 4670 & -- & 2.54 & 1.17 & 233 & 0.079 & 0.117 & 0.067 & $<$ -0.05 \\
		4843 & 11.540 & 1.353 & 0.035 & 4358 & 60 & 1.69 & 1.28 & 240 & -0.006 & 0.089 & 0.063 & $<$ -0.33 \\
		4909 & 12.992 & 1.204 & 0.009 & 4613 & 18 & 2.65 & 1.16 & 246 & -0.128 & 0.131 & 0.076 & $<$ 0.28 \\
		5027 & 11.944 & 1.120 & 0.024 & 4771 & 49 & 2.00 & 1.24 & 318 & -0.051 & -- & -- & 2.04 $\pm$ 0.02\\
		5048 & 13.878 & 1.095 & 0.010 & 4820 & 20 & 3.11 & 1.10 & 126 & 0.021 & 0.200 & 0.116 & $<$ 0.58 \\
		5085 & 12.367 & 1.157 & 0.035 & 4700 & 71 & 2.28 & 1.20 & 84 & 0.234 & 0.146 & 0.073 & $<$ 0.45 \\
		5133 & 14.090 & 1.052 & 0.026 & 4908 & 57 & 3.22 & 1.08 & 99 & 0.240 & 0.171 & 0.121 & $<$ 0.65 \\
		5438 & 13.657 & 1.116 & 0.011 & 4778 & 23 & 3.00 & 1.11 & 124 & 0.082 & 0.144 & 0.083 & $<$ 0.53 \\
		5597 & 12.902 & 1.190 & 0.053 & 4639 & 104 & 2.60 & 1.16 & 230 & 0.010 & 0.112 & 0.056 & $<$ 0.15 \\
		5835 & 12.732 & 1.271 & 0.039 & 4496 & 70 & 2.50 & 1.18 & 119 & -0.017 & 0.153 & 0.088 & $<$ -0.60 \\
		5855 & 13.370 & 1.162 & 0.004 & 4690 & 8 & 2.86 & 1.13 & 114 & 0.120 & 0.134 & 0.067 & $<$ 0.30 \\
		5894 & 14.066 & 1.078 & 0.020 & 4854 & 44 & 3.20 & 1.08 & 142 & 0.038 & 0.155 & 0.078 & $<$ 0.58 \\
		6188 & 13.395 & 1.168 & 0.018 & 4679 & 36 & 2.87 & 1.13 & 173 & 0.134 & 0.078 & 0.039 & $<$ 0.15 \\ 
		6353 & 12.515 & 1.257 & 0.035 & 4520 & 64 & 2.38 & 1.19 & 245 & -0.015 & 0.132 & 0.066 & 0.60 $\pm$ 0.02 \\
		6602 & 12.707 & 1.330 & 0.015 & 4396 & 25 & 2.49 & 1.18 & 208 & 0.078 & 0.143 & 0.072 & $<$ -0.45 \\
		6982 & 12.433 & 1.177 & 0.014 & 4663 & 28 & 2.32 & 1.20 & 265 & -0.172 & 0.045 & 0.032 & $<$ 0.00 \\
		8129 & 13.485 & 1.137 & 0.032 & 4738 & 66 & 2.92 & 1.12 & 151 & 0.245 & 0.263 & 0.186 & $<$ 0.20 \\
		9159 & 12.488 & 1.192 & 0.033 & 4635 & 65 & 2.36 & 1.19 & 124 & 0.252 & 0.097 & 0.056 & $<$ 0.00 \\
		\hline
	\end{tabular}
		\begin{tablenotes}
		\item[1] Average $V$ magnitude, see text.
		\item[2] Average $B-V$ using up to 10 colors, converted to $B-V$. Column 4 shows the standard deviation of the average $B-V$, `--' means the star only has a B and a V magnitude.
		\item[3] $T_{\rm{eff}}$ calculated from the relationship in Ramirez \& Melendez\citet{Ramirez05}; $\sigma(T_{\rm{eff}})$ is derived based on error propagation from $\sigma\ (B-V)$.
		\item[4] Poisson-based signal-to-noise ratio (SNR).
		\item[5] Iron abundance, standard deviation ($\sigma$), and standard deviation of the mean ($\sigma_{\mu}$) for individual stars by averaging all the qualifying Fe I lines.
		\item[6] Li abundance A(Li) or 3$\sigma$ upper limit A(Li). For the four stars with detectable Li, 1$\sigma$ error is calculated using the relationship from Deliyannis et al.\citet{Deliyannis93a}, where here the effective pixel size d = 0.205 \AA.
	\end{tablenotes}
	\end{threeparttable}
\end{table*}

\begin{figure*}
	\centering
	\includegraphics[width=1.0\textwidth]{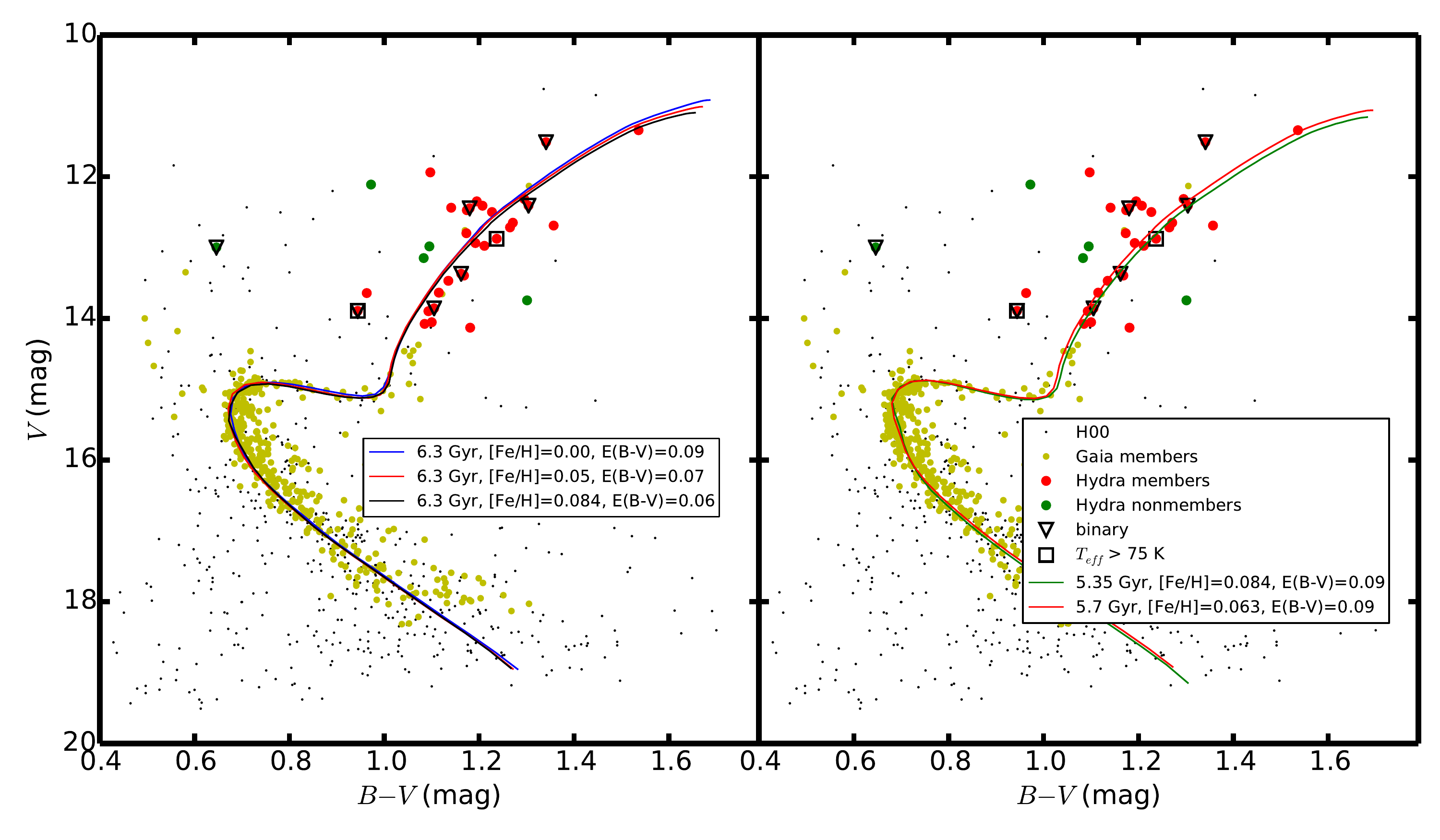}
	\caption{CMD in the central region (Radius $<$ 12') of NGC 188. The black points are stars with H00 photometry; the yellow points are Gaia members; the red points are Hydra members; the green points are Hydra nonmembers. Binaries are shown in downward open triangles, and two stars with $\sigma_{Teff} >$ 75 K are shown in open squares. Yale-Yonsei isochrones are shown with $(m - M)_V = 11.40$ mag in the left panel. In the right panel, the [Fe/H] = 0.084 dex isochrone has $(m - M)_V = 11.55$ mag; the [Fe/H] = 0.063 dex isochrone has $(m - M)_V = 11.48$ mag.}
	\label{fig:Isochrone}
\end{figure*}

We derive $T_{\rm{eff}}$ by employing the $T_{\rm{eff}}$--($B-V$) relationship for red giant stars from Ramirez \& Melendez\citet{Ramirez05}.  To begin with, we assume [Fe/H] = 0.00 dex, which is in the middle of the range of published values (see Section \ref{sec:metal}), and adopt interstellar reddening $E(B-V)$ = 0.09 $\pm$ 0.02 mag from S99. The final (after iteration, see below) $T_{\rm{eff}}$ and $\sigma\ (T_{\rm{eff}})$ (standard deviation based on error propagation from $\sigma\ (B-V)$) are shown in Table \ref{tab:atmosphere}. 

Figure \ref{fig:Isochrone}, left panel, shows the H00 photometry. A 6.3 Gyr Yale-Yonsei\citet[Demarque et al.][hereafter $Y^2$]{Demarque04} isochrone fits well the MS just below the turnoff, the turnoff, and subgiants just beyond the turnoff, so we use this isochrone to derive surface gravity (log $g$) at the $V$ of the stars. We derive microturbulence ($\xi$) using the relation $\xi\ =\ 1.5 - 0.13\ \rm{log}\ {\it g}\ km\ s^{-1}$ from Carretta et al.\citet{Carretta04} for red giant stars. Table \ref{tab:atmosphere} includes our final (after iteration, see below) values for log $g$ and $\xi$.  Fortunately, only a few stars have $\sigma(T_{\rm{eff}}) >$ 75 K. 

\subsection{Metallicity} \label{sec:metal}

The metallicity of NGC 188 is not only important for our study, but since NGC 188 is a very old open cluster, the metallicity is also of importance to studies of Milky Way formation and evolution.  Unfortunately, even recent high resolution spectroscopic studies are discrepant with one another (see discussion below). We aim to shed some light on this issue, using the largest sample of cluster stars to date.

\begin{figure}
	\centering
	\includegraphics[width=0.45\textwidth]{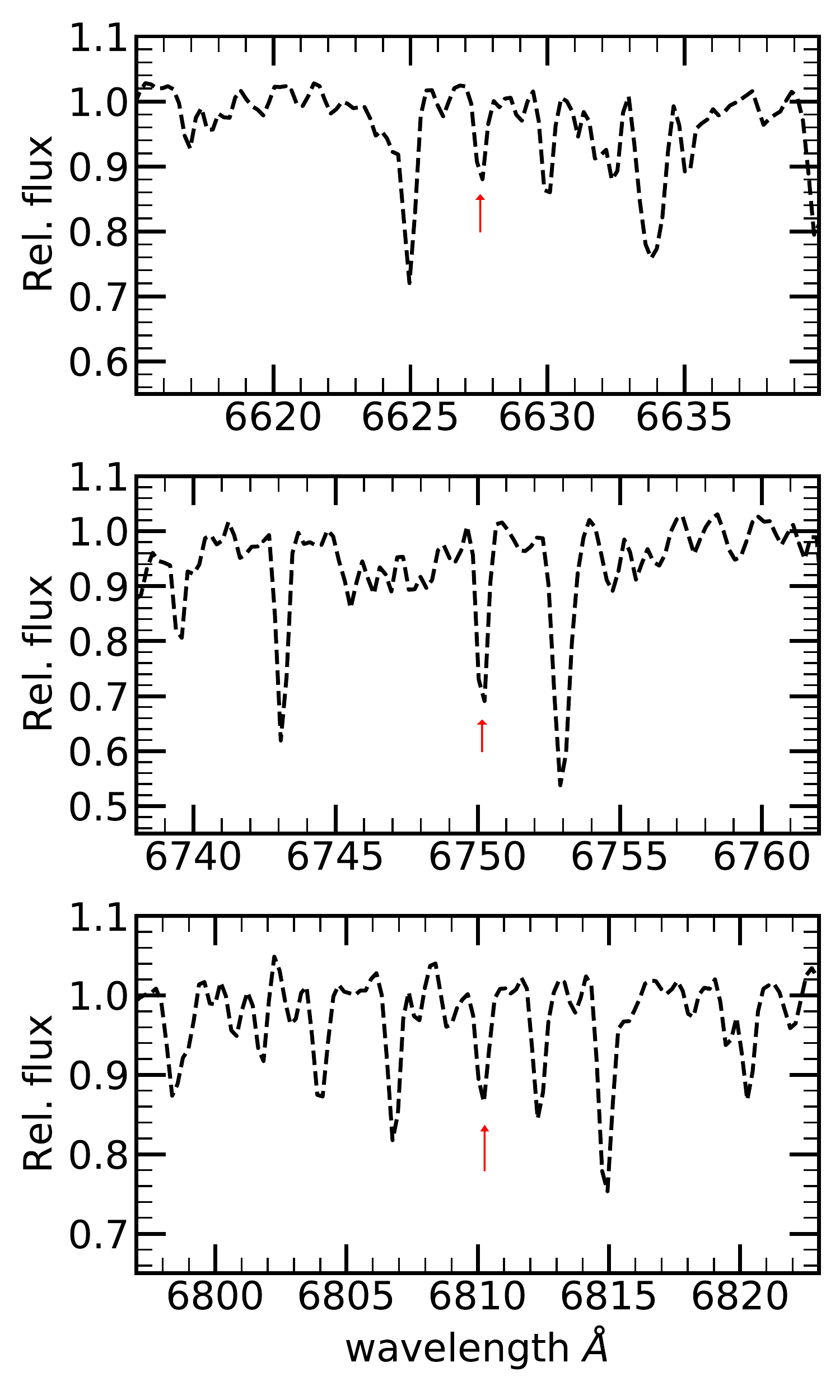}
	\caption{Examples of our selected Fe I lines at 6627.544 $\AA$ (star 3271); 6750.152 $\AA$, and 6810.262 $\AA$ (star 4756).}
	\label{fig:iso_lines_eg}
\end{figure}

We selected 7 isolated Fe I lines that showed no significant blending out to at least 0.5 \AA\ on either side of the line center in the Arcturus spectrum (R$\sim$150000, Hinkle et al.\citealt{Hinkle00}), and measured the equivalent width of each line for the 29 members of NGC 188. Table \ref{tab:FeI_param} shows the wavelength $\lambda$ (\AA), Excitation Potential (eV), and log (gf) values of these lines. Figure \ref{fig:iso_lines_eg} shows a few examples of the lines that were used. Using the Kurucz\citet{Kurucz92} model atmosphere grids with [Fe/H] = 0.0 dex, we interpolated to construct model atmospheres matching the $T_{\rm{eff}}$, log $g$, and $\xi$ values for each star. Then, we performed LTE analysis for each Fe I line for each star using the {\it abfind} task of MOOG\citet[Sneden][]{Sneden73} to derive A(Fe). We measured the Poisson-based signal-to-noise ratio (SNR) per pixel in a relatively line-free region near the Li 6707.8 \AA\ line (see Table \ref{tab:atmosphere}).

\begin{table}
	\renewcommand{\arraystretch}{1.3}
	\caption{Fe I line parameters}
	\label{tab:FeI_param}
	\centering
	\begin{tabular}{ccc}
		\hline
		$\lambda$ & Excitation & log (gf) \\
		\AA\ & Potential (eV) &  \\
		\hline
		6546.24 & 2.76 & -1.90 \\ 
		\hline
		6597.56 & 4.80 & -1.04 \\
		\hline 
		6627.54 & 4.55 & -1.57 \\
		\hline 
		6710.32 & 1.49 & -4.77 \\ 
		\hline 
		6726.67 & 4.61 & -1.12 \\
		\hline 
		6750.15 & 2.42 & -2.48 \\
		\hline 
		6810.27 & 4.61 & -1.12 \\ 
		\hline
	\end{tabular}
\end{table}

\begin{figure*}
	\centering
	\includegraphics[width=0.7\textwidth]{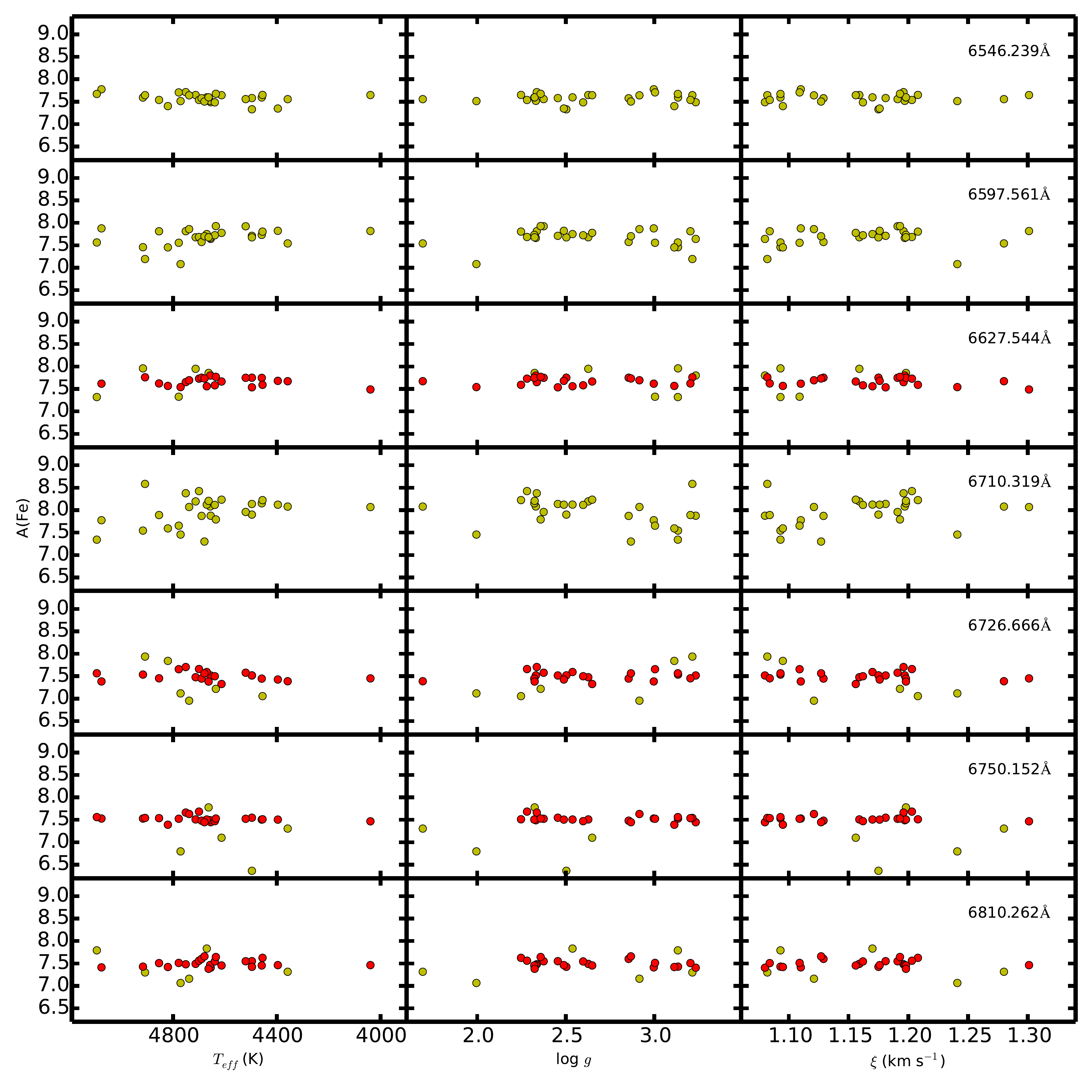}
	\caption{The dependence of A(Fe) on $T_{\rm{eff}}$ (K), log $g$, and $\xi$ for different $\lambda$. Colored circles show iron abundances for each Fe I lines for each apertures. Red circles are measurements that are kept for the final cluster average A(Fe) calculation; yellow circles are outliers rejected from the calculation.}
	\label{fig:abun_line}
\end{figure*}

\begin{figure}
	\centering
	\includegraphics[width=0.5\textwidth]{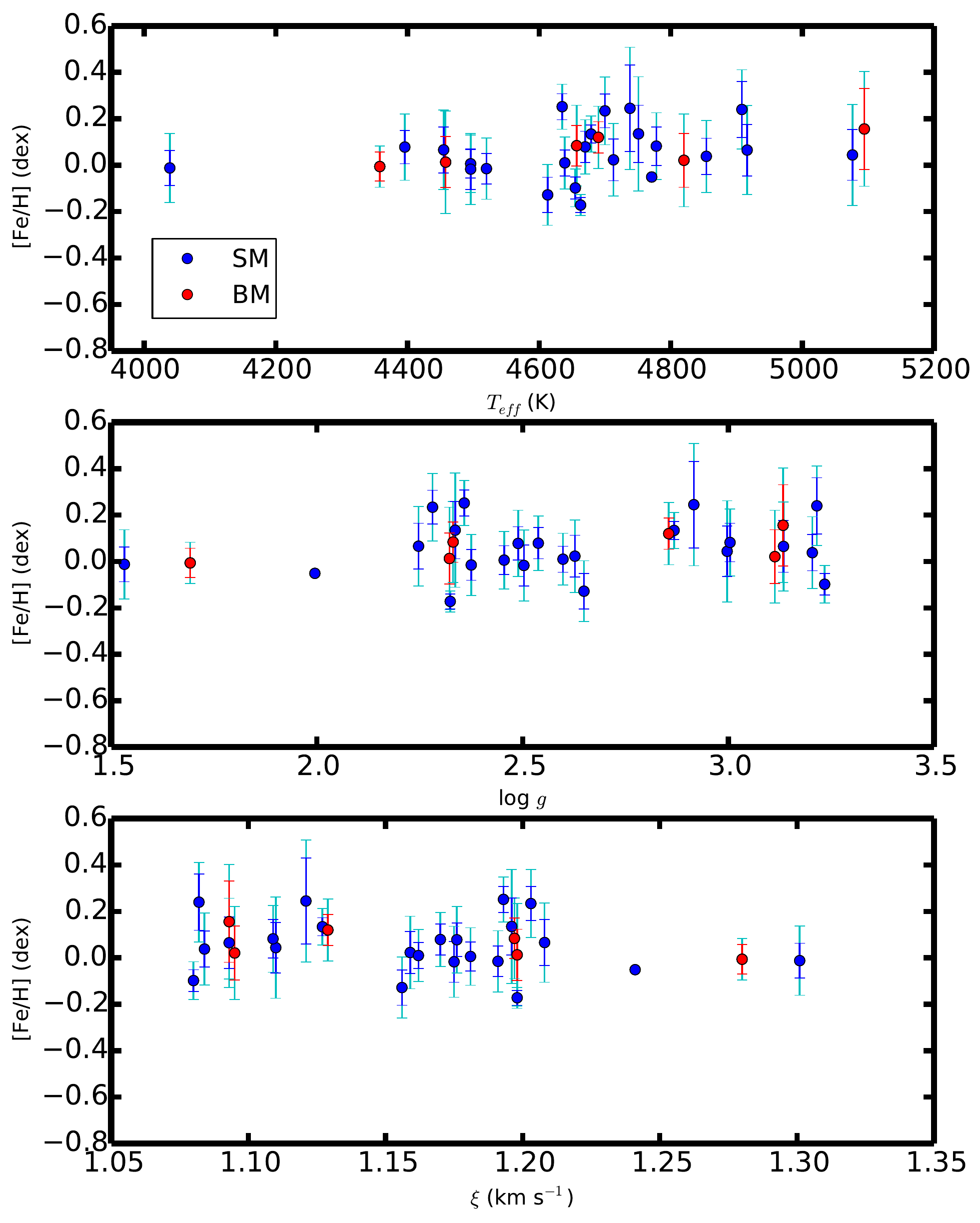}
	\caption{Average [Fe/H] of the 29 member stars, cyan error bars are standard deviation $\sigma$, blue error bars are standard deviation of the mean $\sigma_{\mu}$. Red disks are BM, blue disks are SM.}
	\label{fig:abun_star_sel}
\end{figure}

Figure \ref{fig:abun_line} shows the final (after iteration, see below) A(Fe) derived for each line for each star as a function of the stellar $T_{\rm{eff}}$, log $g$, and $\xi$. We excluded (yellow disks) some entire lines that show dependence on $T_{\rm{eff}}$, log $g$, or $\xi$ ($\lambda$ = 6546.239 \AA) and/or excessive scatter ($\lambda$ = 6597.561 \AA, 6710.319 \AA) and, for the remaining lines, we excluded those from individual stars that show excessive scatter. To derive [Fe/H] we have employed the method of ``solar gf-values'', which we also used in Sun et al.\citet{Sun20}, calculating [Fe/H] strictly relative to the Sun. In particular, for each line and for each aperture, we derive solar A(Fe) by measuring the same Fe I lines using sky spectra taken from the same configuration. For each line, we then subtracted the solar A(Fe) from the stellar A(Fe), using the solar spectrum from the same fiber as the stellar spectrum to help maximize consistency. For each star, we used the selected lines to calculate an average [Fe/H] in linear (not log) space, weighted linearly (not quadratically) by the 1$\sigma$ error of each line (see also C17).  This error is calculated based on the SNR and FWHM of the Fe I line, using the relationship from Deliyannis et al.\citet{Deliyannis93a}. We also calculated $\sigma$ and $\sigma_{\mu}$ for each star. We show the final (after iteration, see below) results in Table \ref{tab:atmosphere} and Figure \ref{fig:abun_star_sel}.

For the {\it cluster}, we used all the surviving individual Fe I lines to derive a cluster average (again weighted, in linear space) of $[Fe/H]_{188}$ = +0.084 $\pm$ 0.016 dex ($\sigma_{\mu}$, and $\sigma$ = 0.157 dex).  This value is very close to the unweighted average $[Fe/H]_{188}$ = +0.087 $\pm$ 0.020 dex ($\sigma_{\mu}$, and $\sigma$ = 0.193 dex). If instead we keep only the 23 SM, we derive a weighted cluster average $[Fe/H]_{188}$ = +0.083 $\pm$ 0.019 dex ($\sigma_{\mu}$, and $\sigma$ = 0.164 dex), and an unweighted cluster average of $[Fe/H]_{188}$ = +0.088 $\pm$ 0.023 dex ($\sigma_{\mu}$, and $\sigma$ = 0.197 dex). Although in principle one might be concerned that the secondary in each BM might contaminate the spectrum sufficiently so as to affect the measured equivalent widths, the above averages are indistinguishable whether they include the BM or not, and Figure \ref{fig:abun_star_sel} also shows that the BM have indistinguishable [Fe/H] compared to SM stars.

These values for the cluster average [Fe/H] are sufficiently different from our assumed value of 0.00 dex for the isochrone that we decided to fine-tune by iterating. Figure \ref{fig:Isochrone}, right panel, shows that a similar fit can be provided with a 5.35 Gyr isochrone that has [Fe/H] = +0.084 dex, the same $E(B-V)$ = 0.09 mag, and $(m - M)_V = 11.55$ mag. We now repeated the above analysis of determining log $g$, $\xi$, and abundances, and found new weighted and unweighted averages for [Fe/H].  This process was iterated a few more times until the input and output [Fe/H] converged. Our final weighted average [Fe/H] for the cluster is +0.064 $\pm$ 0.018 dex ($\sigma_{\mu}$, and $\sigma$ = 0.177 dex), very close to the unweighted value of +0.069 $\pm$ 0.019 dex ($\sigma_{\mu}$, and $\sigma$ = 0.186 dex); excluding BM we find again nearly identical results. We stress that Table \ref{tab:atmosphere} and Figures \ref{fig:abun_line} and \ref{fig:abun_star_sel} show these final abundances. The trends discussed above in these figures are nearly identical as the original ones.

Unlike the original isochrone (Figure \ref{fig:Isochrone}, left panel), the iterated younger isochrones (5.35 Gyr, [Fe/H] = +0.084 dex; 5.7 Gyr, [Fe/H] = +0.063 dex, right panel) show a slight kink at the turnoff that does not seem to be present in the data. One way to avoid the kink is to keep the older age of 6.3 Gyr and allow the reddening to vary. A very similar fit as that with the original isochrone, and no kink at the turnoff, can be achieved with [Fe/H] = +0.084 dex, $E(B-V)$ = 0.06 mag and the same distance modulus (left panel). Anticipating a slightly lower final [Fe/H], we use an isochrone with [Fe/H] = +0.05 dex, which also provides a nearly identical fit with $E(B-V)$ = 0.07 mag (left panel), which is at the low end of the error range in reddening. Repeating the analysis now results in an average [Fe/H] for the cluster +0.060 $\pm$ 0.017 dex ($\sigma_{\mu}$, and $\sigma$ = 0.162 dex), weighted, or +0.064 $\pm$ 0.018 dex ($\sigma_{\mu}$, and $\sigma$ = 0.175 dex), unweighted.  Encouragingly, this is very close to the final results discussed above, and suggests that systematic errors potentially introduced by these slightly different choices in parameters have little effect on the results.

An independent analysis of the solar spectra using laboratory gf-values gives an average A(Fe)$_{\odot}$ = 7.520 $\pm$ 0.015 dex ($\sigma_{\mu}$, $\sigma$ = 0.164 dex), which is in excellent agreement with previous studies (e.g. Lodders et al.\citealt{Lodders09}). Table \ref{tab:sys_error} shows how [Fe/H] depends on $B-V$, $T_{\rm{eff}}$, log $g$, and $\xi$ at three different $T_{\rm{eff}}$ and for the whole cluster.

\begin{table*}
	\caption{Estimation of systematic errors on [Fe/H]}
	\label{tab:sys_error}
	\begin{threeparttable}
	\begin{tabular}{ccccc}
	\hline
	{Parameter Changes}$^1$ & {4000 K}$^2$ & {4500 K}$^2$ & {5000 K}$^2$ & NGC 188 cluster$^3$ \\
	
	\hline
	$\Delta(E(B-V))$ = +0.01 mag& 0.035 & -0.038 & -0.017 & -0.035 \\
	$\Delta(T_{eff})$ = +100 K& -0.060 & -0.011 & 0.058 & 0.004 \\
	$\Delta$(log {\it g}) = +0.2 & 0.063 & 0.010 & -0.003 & 0.005\\
	$\Delta(\xi)$ = +0.2 km s$^{-1}$& 0.013 & -0.048 & -0.015 & -0.043 \\
	$\Delta_{comb}(\Delta(E(B-V$)) = +0.01 mag) & 0.023 & -0.022 & -0.035 & -0.026 \\ 
	\hline
	\end{tabular}
	\begin{tablenotes}
	\item[1] In the first four rows, we show changes in [Fe/H] when changing one parameter at a time and keeping the other parameters fixed. The last row shows changes in [Fe/H] when $\Delta(E(B-V$)) = +0.01 mag and the other three parameters are also changed accordingly.
	\item[2] [Fe/H] change for a star at $T_{\rm{eff}}$ = 4000 K, 4500 K, 5000 K.
	\item[3] [Fe/H] change for the whole NGC 188 cluster following the above procedure.
\end{tablenotes}
\end{threeparttable}
\end{table*}

There have been several high resolution spectroscopic studies during the past decade, but only two before that (summarized in Table \ref{tab:summary_metal}). The results span the range from Fe/H] = -0.12 to +0.14 dex. Note that the two reduction methods used by Casamiquela et al.\citet{Casamiquela17} on the same data gave results that differed by 0.07 dex.  In fact, it seems the high-R studies in Table \ref{tab:summary_metal} cluster either near solar or 0.1 dex higher and, taking the stated errors at face value, they do so in a discrepant way. For example, Donor et al.\citet{Donor18} and Jacobson \& Friel\citet{Jacobson13} are discrepant with Casamiquela et al.\citet{Casamiquela17}, Jacobson et al.\citet{Jacobson11}, and Randich et al.\citet{Randich03}; this is true even when considering only the two more recent studies of Donor et al.\citet{Donor18} and both analyses of Casamiquela et al.\citet{Casamiquela17}. Our own result is between the two clustered sets of results. Comparing to the only high-R study with a similar number of stars, namely Jacobson et al.\citet{Jacobson11}, for 18 stars in common we find average differences in $T_{\rm{eff}}$, log $g$, and $\xi$ of 11 K, 0.17, and 0.34 km s$^{-1}$, respectively, in the sense theirs minus ours. The differences in $T_{\rm{eff}}$ and log $g$ have a negligible effect ($<$ 0.005 dex) on the derived cluster [Fe/H] but adopting the average difference in $\xi$ would decrease our cluster [Fe/H] by 0.073 dex, accounting for the difference in [Fe/H] between our studies of 0.09 $\pm$ 0.04 dex.

We point out that abundances from medium-resolution studies span a similar range in [Fe/H] as from the high resolution studies, though with larger errors and thus there is no issue of discrepant results, see Table \ref{tab:summary_metal}. Photometrically-derived metallicities span a larger range, and a number of them have been summarized in Sun\citet{Sun21}.

\begin{table*}
	\caption{Summary of past spectroscopic metallicity studies}
	\label{tab:summary_metal}
	\begin{threeparttable}
	\resizebox{2.\columnwidth}{!}{%
	\begin{tabular}{cccccc}
	
	\hline
	{[Fe/H]$^{1}$ (dex)} & {number}$^{1}$ & {star type}$^{2}$ & {$\lambda$}$^{1}$ & R$^{1}$ & References and comments \\
	
	\hline
	\multicolumn{6}{c}{\textbf{High resolution spectroscopic studies}} \\
	+0.064 $\pm$ 0.018 & 29 & RC, RGB & 6428--6847 \AA & 17000 & this study (weighted average is preferred); WIYN/Hydra \\ 
	+0.14 $\pm$ 0.03 & 13 & ? & 1.51--1.70 $\mu$m & 22500 & Donor et al.\citet{Donor18}; APOGEE \\
	+0.02 $\pm$ 0.02 & 6 & RC & 4000--9000 \AA & $>$ 65000 & Casamiquela et al.\citet{Casamiquela17}; GALA reduction \\
	-0.05 $\pm$ 0.05 & 6 & RC & 4000--9000 \AA & $>$ 65000 & Casamiquela et al.\citet{Casamiquela17}; iSpec reduction\\ 
	+0.11 $\pm$ 0.04 & 4 & -- & -- & -- & Netopil et al.\citet{Netopil16}; homogenized reanalysis of previous studies, see also Heiter et al.\citet{Heiter14} \\ 
	+0.12 $\pm$ 0.04 & 4 & RC/RGB & 4800--8300 \AA & 28000 & Jacobson \& Friel\citet{Jacobson13}; KPNO 4m, see also Friel et al.\citet{Friel10} \\ 
	-0.03 $\pm$ 0.04 & 27 & evolved & 6130--6430 \AA & 18000 & Jacobson et al.\citet{Jacobson11}; WIYN/Hydra \\
	+0.01 $\pm$ 0.08 & 5 & dwarf & 5500--11000 \AA & 35000-57000 & Randich et al.\citet{Randich03} \\
	-0.12 $\pm$ 0.16 & 7 & turnoff & 5100--6900 \AA & 30000 & Hobbs et al.\citet{Hobbs90} \\	
	\multicolumn{6}{c}{\textbf{Medium/low resolution spectroscopic studies}} \\
	+0.08 $\pm$ 0.05 & 14 & K giants & 3700--6000 \AA & 1000 & Worthey \& Jowett\citet{Worthey03} \\
	+0.10 $\pm$ 0.09 & 21 & ? & 4000--6000 \AA & 1250 & Friel et al.\citet{Friel02} \\
	+0.05 $\pm$ 0.38 & 4 & giants & 4700-5650 \AA & 2060, 1200 & Thogersen et al.\citet[][T93]{Thogersen93}\\
	-0.026 $\pm$ 0.012 & 10 & giants & 4700-5650 \AA & 2060, 1200 & Twarog et al.\citet{Twarog97}; reanalyzed T93 together with DDO photometry of Piatti et al.\citet[][P95]{Piatti95}\\
	\hline
	\end{tabular}
}
	\begin{tablenotes}
	\item{1} [Fe/H], number of stars, wavelength range, and spectroscopic resolution of the given study.
	\item{2} The kind of stars observed, ``RC'' stands for ``Red Clump'', ``RGB'' stands for ``Red Giant Branch'', ``AGB'' stands for ``Asymptotic Giant Branch''.
	\end{tablenotes}
	\end{threeparttable}
\end{table*}

\section{Lithium Abundances}  \label{sec:li_abun}

An Fe I line at 6707.43 \AA\ blends with the Li 6707.8 \AA\ line and must sometimes be taken into account (Thorburn et al.\citealt{Thorburn93}; Steinhauer \& Deliyannis\citealt{Steinhauer04}; C17). This is especially true for cooler stars where the Fe line becomes stronger and the Li line becomes weaker (due to stellar depletion). Here, using the {\it synth} task in MOOG, we generated synthetic spectra in the Li region (6700-6715 \AA) using the line-list from C17. We assume any contribution from the $^6$Li isotope is negligible, for the following reasons:

Of the two stable Li isotopes, only 7.5\% of the meteoritic Li is in the form of $^6$Li, and the remainder is $^7$Li\citet[Lodders et al.][]{Lodders09}; therefore, the Sun itself likely formed with this ratio. It is not unreasonable to assume that other solar-metallicity stars also formed with a similar ratio. From the final isochrone fit of Section \ref{sec:iron_abun}, current (normal) giants in NGC 188 have masses of $\sim$1.14 $M_{\odot}$, whereas current (normal) turnoff stars have masses $\sim$1.09 $M_{\odot}$ and $T_{\rm{eff}} \sim 6000\ K$\citet[Hobbs et al.][]{Hobbs90}. $^6$Li is more fragile than $^7$Li and standard models alone deplete $^6$Li in corresponding stars during the pre-MS by factors of several orders of magnitude\citet[Proffitt \& Michaud][]{Proffitt89}, and rotational mixing depletes both isotopes ($^6$Li more so, Pinsonneault et al.\citealt{Pinsonneault90}) during the MS. Finally, subgiant dilution is about a factor of 2 more severe for $^6$Li as compared to $^7$Li, and there may be additional rotational mixing on the subgiant branch as evidenced by correlated Li and Be depletion in M67 subgiants\citet[Boesgaard et al.][]{Boesgaard20}.  We thus expect that $^6$Li is negligible in giants of NGC 188.

Our data provide us with the opportunity to refine the line list further. The possibility has been raised that some very weak lines exist right on top of the Li I doublet (see detailed discussion in King et al.\citealt{King97}). These are expected to be extremely weak ($<$ 1 m\AA) and thus not affect the vast majority of A(Li) determinations. However, in stars with very low A(Li), limited knowledge of these lines may hinder accurate determination of A(Li), making it difficult to make progress on questions such as how does A(Li) decline in Hyades dwarfs cooler than 5100 K (Thorburn et al.\citealt{Thorburn93}; C17) and in other open clusters of similar age or older, and how deep is the Li Dip? Answers to these questions would help elucidate the physical processes that deplete surface Li abundances. Even in the Sun, the Li equivalent width is a few m\AA, so improved knowledge of these lines could lead to an improved A(Li) in the Sun and other similar stars. King et al.\citet{King97} improved the line list (used by our group ever since, including in C17) in the Li I region using the Sun, 16 Cyg A and B, and $\alpha$ Cen A and B. Since candidate lines get stronger with lower $T_{\rm{eff}}$ and unsuspected ionized lines may get stronger with lower gravity, we use the brightest, coolest giants that are not cool enough to show evidence of molecular lines to further calibrate the line list.  

\begin{figure*}
	\centering
	\includegraphics[trim=0cm 8cm 0cm 8cm,width=1.0\textwidth]{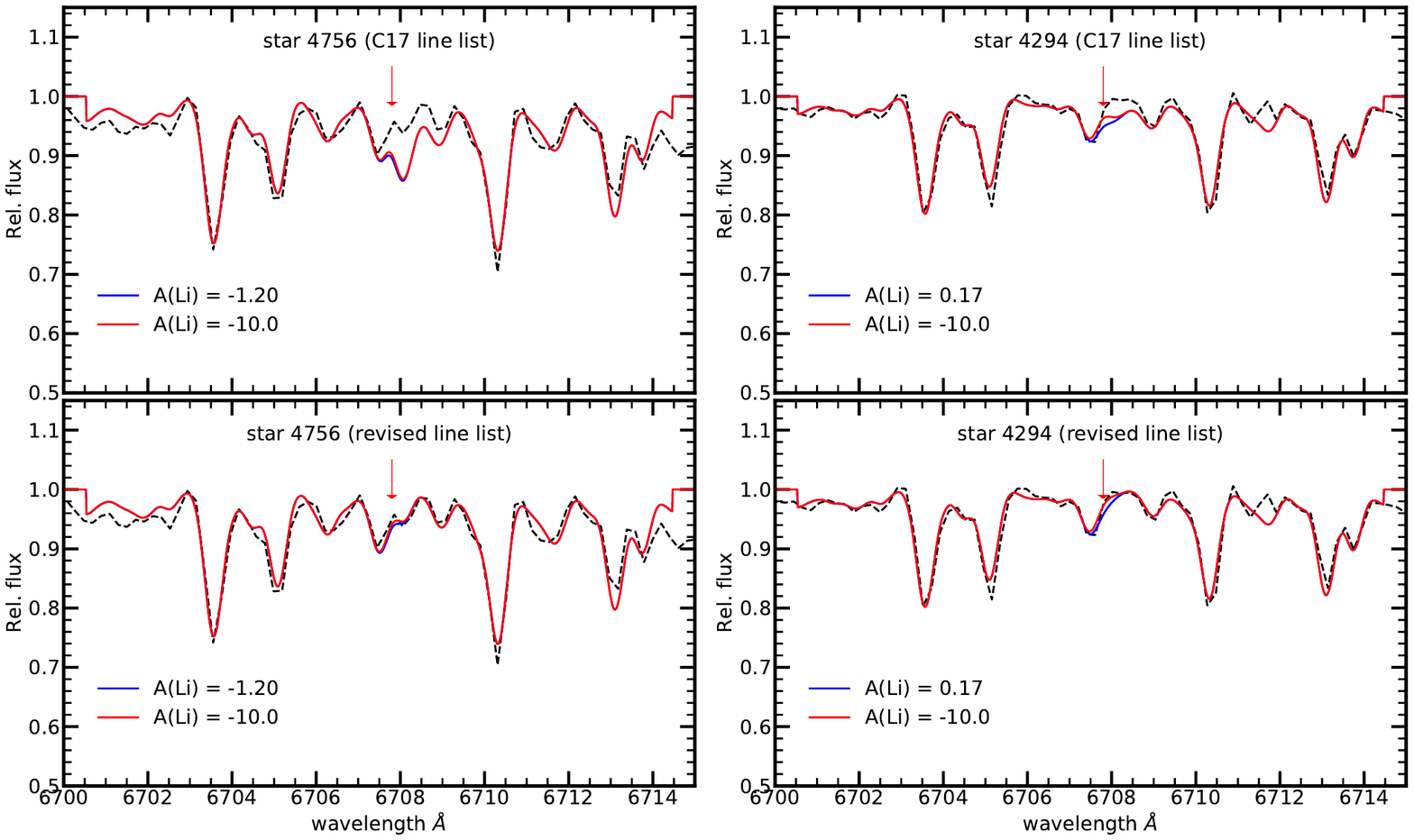}
	\caption{Comparison of the syntheses of star 4756 using C17 line list (top panel) and our revised line list (bottom panel). The red line shows a synthesis with no Li, and the blue line shows a synthesis using a 3$\sigma$ upper limit A(Li).}
	\label{fig:star4756_4294}
\end{figure*}

The top left panel of Figure \ref{fig:star4756_4294} show a synthesis using the C17 line list against our spectrum of star 4756, the coolest and brightest giant SM ($T_{\rm{eff}} = 4039\ K$) for which many lines are stronger than in any other star. It is clear that the synthesis has too much absorption at the position of the Li line (vertical arrow). The conservative approach to adjusting the line list is to assume that the star is devoid of Li, and thus that the observed absorption is due to non-Li lines at the same position. If some of the absorption was due to Li, then even better constraints could be derived for the other lines, but we cannot know a priori whether any Li is present. Even so, the constraints on the line list are being improved. We adjusted {\it gf} values of lines in the 6707.0 -- 6709.2 \AA\ region using star 4756 and the high quality and high resolution (R$\sim$47000) spectrum of the metal-rich giant $\mu$ Leo from the library of the Gaia benchmark stars  (\url{https://www.blancocuaresma.com/s/benchmarkstars}), The latter resolves some lines that are blended in our Hydra spectra that can now be matched more clearly and explicitly as we revised the line list. We used our own parameters for this star, which turned out to agree extremely well with the Gaia values. Finally, fine-tuning with several stars at various temperatures ensured appropriate temperature dependence of the final line list. The bottom left panel of Figure \ref{fig:star4756_4294} shows that the resulting synthesis (red) matches the observed spectrum better than the C17 line list. The right panels of Figure \ref{fig:star4756_4294} show corresponding syntheses for star 4294, which is almost 700 K hotter than star 4756. The blue curves show synthetic Li with an abundance corresponding to a detection in that spectrum at the $3\sigma$ level. We use this revised line list henceforth.

Debate continues about the possible importance of non-LTE effects to the Li line. Carlsson et al.\citet{Carlsson94} suggest any such effects would be $<$ 0.1-0.2 dex and similar for stars in a similar evolutionary state, and thus unimportant for our purposes. Note also that spot coverage may possibly result in spuriously large A(Li) in rapidly rotating cooler dwarfs\citet[Jeffries et al.][]{Jeffries21}, but our giants are rotating slowly.

\begin{figure}
	\centering
	\includegraphics[width=0.45\textwidth]{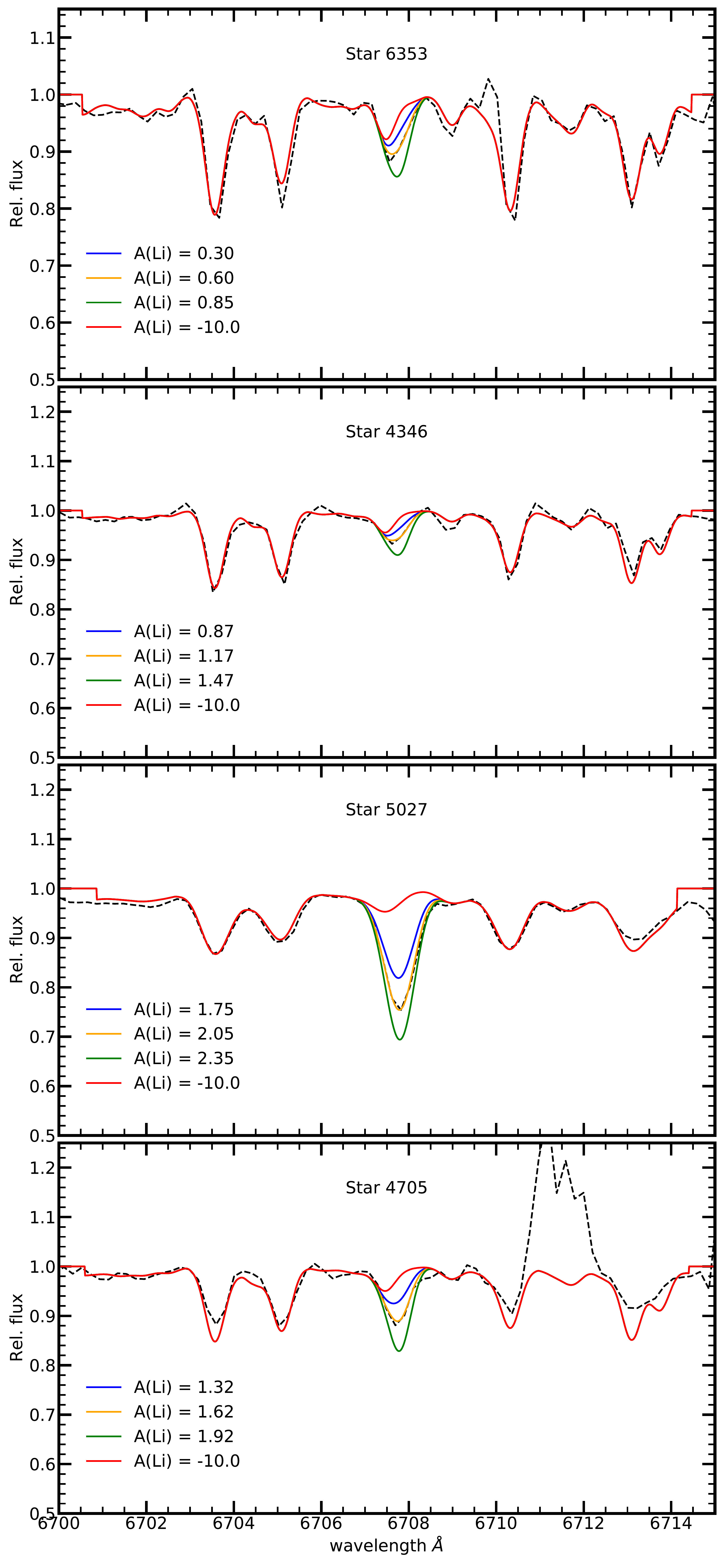}
	\caption{Synthetic spectra for stars 6353, 4346, 5027, 4705 in the Li 6707.8 \AA\ region. The red line shows the no Li condition (A(Li) = -10.0 dex). The green and blue lines show the A(Li) that are 2 times larger and smaller, respectively, than the best fit A(Li).}
	\label{fig:comb_ALi}
\end{figure}

Figure \ref{fig:comb_ALi} shows spectra and syntheses for the four stars for which we have confident detections of Li. Synthetic spectra are shown for no Li (A(Li) = -10.0 dex, red line), the best fit (orange line), and twice/half of the A(Li) for the best fit (green/blue). The resulting best fits have A(Li) = 2.04 dex for star 5027, 0.60 dex for star 6353, 1.17 dex for star 4346, and 1.65 dex for star 4705.

\begin{figure*}
	\centering
	\includegraphics[width=0.95\textwidth]{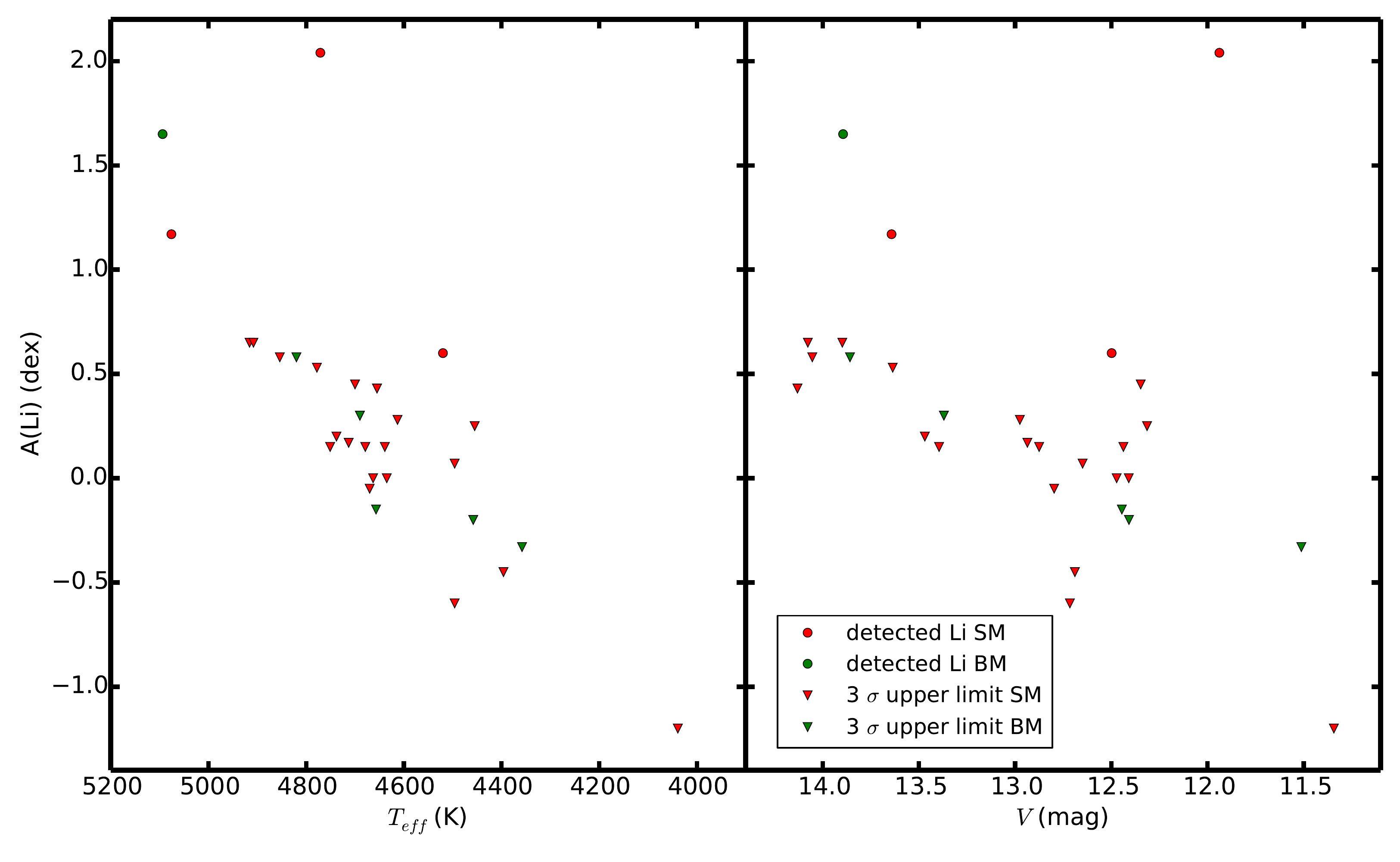}
	\caption{A(Li) as a function of $T_{\rm eff}$ (left panel) and $V$ magnitude (from H00, right panel).}
	\label{fig:ALi_teff}
\end{figure*}

\begin{figure*}
	\centering
	\includegraphics[width=1.0\textwidth]{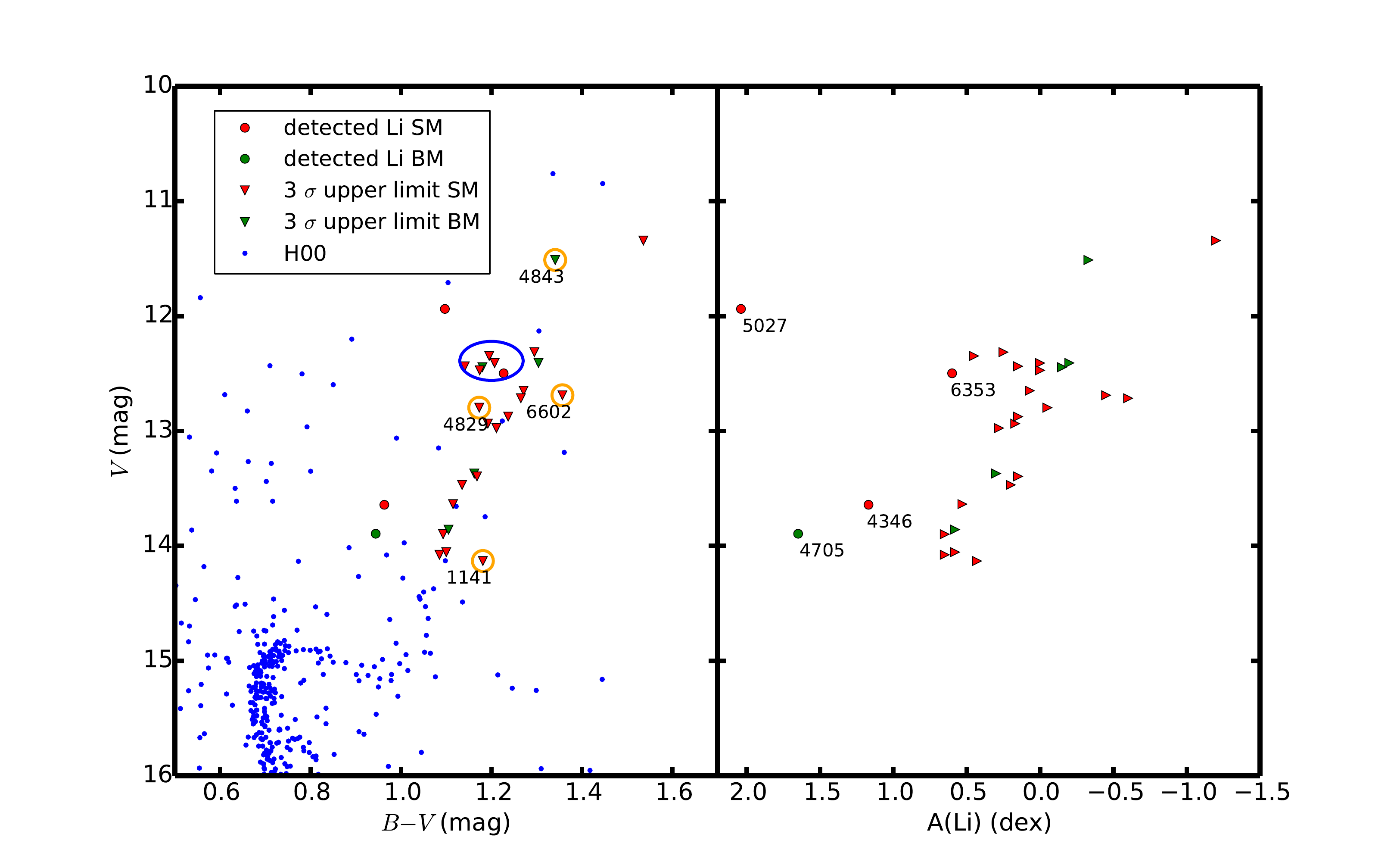}
	\caption{A(Li) of member stars on the CMD. In the left panel, the blue circle highlights the apparent red clump stars, while the orange circles highlight some other stars that lie slightly off of the well-defined fiducial sequence.}
	\label{fig:CMD_ALi}
\end{figure*}

For the remaining 25 stars we report 3$\sigma$ upper limits calculated using the relationship from Deliyannis et al.\citet{Deliyannis93a}. Table \ref{tab:atmosphere} shows the Li abundances. Figure \ref{fig:ALi_teff} shows A(Li) plotted against $T_{\rm eff}$, and $V$ magnitude. Figure \ref{fig:CMD_ALi} shows the location of the 29 member stars on the CMD and their A(Li).

\section{Discussion}  \label{sec:explanation}

\subsection{Using Star Clusters to Define ``Li-rich''}

As noted in the Introduction, a particular advantage of studying Li-rich giants in clusters is that clusters provide an invaluable context for the phenomenon by allowing comparison of Li-rich giants to stars of similar mass, composition, age, and evolutionary state, and by supplying a baseline of unevolved stars at the turnoff to constrain the Li evolution of the progenitors of the Li-rich giants. This level of detail can lead to greater insight than may be available in even very large field star samples. As an example, WOCS 7017, a canonically-defined Li-rich giant in NGC 6819 (Anthony-Twarog et al.\citealt{Anthony13}), has been shown to have a substantially lower mass than the other member giants (Carlberg et al.\citealt{Carlberg15}, Handberg et al.\citealt{Handberg17}), suggesting that the origin of its unusual surface A(Li) may be tied to severe mass loss, possibly related to $^7$Be-transport during a particularly severe He-core flash, and implying that some of the surprisingly low giant masses identified in Deepak et al.\citet{Deepak20} may be real. Another example is W2135 in NGC 2243: the unique properties of this star (enhanced Li with binarity and rapid rotation) compared to other member giants are consistent with mass transfer of Li-rich material from a former AGB star\citet[Anthony-Twarog et al.][]{Anthony20}.

\begin{figure}
\centering
\includegraphics[width=0.45\textwidth]{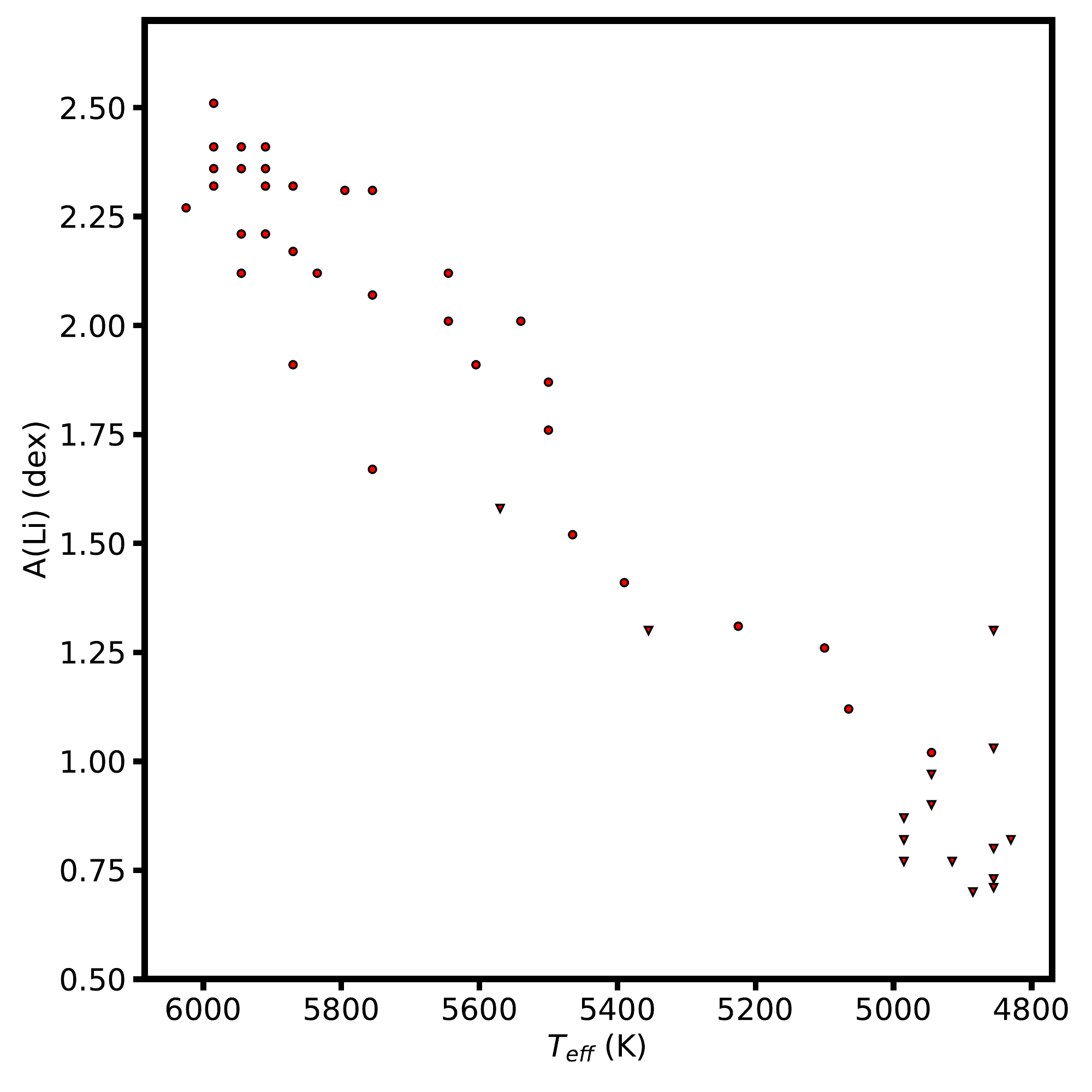}
\caption{Li depletion pattern of subgiants in NGC 188. The red disks are Li detections, the red downward triangles are 3$\sigma$ upper limits.}
\label{fig:subgiantLi}
\end{figure}

The relevant context for NGC 188 goes to the very heart of the canonical definition of ``Li-rich'' for giants as A(Li) $>$ 1.5 dex. While such a definition may be generically adequate for field stars whose Li evolution during the MS is unknown, i.e. consistent with subgiant dilution by 1.8 dex from a maximum initial A(Li) = 3.3 dex (see the discussion in the Introduction), in a star cluster one can constrain {\it much more precisely} what the Li evolution of the progenitors of the Li-rich giants may have been. In the case of NGC 188, A(Li) at the turnoff has been measured at $\sim$ 2.5 dex or slightly lower (Hobbs \& Pilachowski\citealt{Hobbs88}; Randich et al.\citealt{Randich03}), with a steady decline on the subgiant branch to upper limits of $\sim$ 0.8 dex or slightly lower (Deliyannis et al.\citealt{Deliyannis22}, Figure \ref{fig:subgiantLi}). Among our RGs on the well-delineated first-ascent branch and within the RC (Figure \ref{fig:CMD_ALi}), all have upper limits (25 stars) or a detection (1 star) of A(Li) $<$ 0.7 dex (Figure \ref{fig:ALi_teff}). This pattern is similar to what is observed in the 4 Gyr-old M67 \citet[Sills \& Deliyannis][]{Sills00}, though the decline isn't quite as steep. But the similarity to M67 may be quite telling: Li data in M67 subgiants, together with Be data in the same stars, point strongly to the action of rotational mixing as the dominant Li- and Be- depletion mechanism during MS evolution and, together with dilution on the subgiant branch \citet[Boesgaard et al.][]{Boesgaard20}, it is likely the same mechanisms govern the Li depletion in the NGC 188 subgiants. {\it None of the 47 subgiants in NGC 188 deviate from this pattern}, i.e. none are found above this A(Li)-$T_{\rm{eff}}$ pattern and thus none have preserved an unusually high amount of Li (Figure \ref{fig:subgiantLi}). This is not surprising since rotational mixing induced by angular momentum loss depletes Li throughout the Li preservation region (the region in the outer layers that remain sufficiently cool to preserve Li), not just at the surface (Pinsonneault et al.\citealt{Pinsonneault90}; Deliyannis \& Pinsonneault\citealt{Deliyannis97}). So it is not unreasonable to assume that giants which have evolved further compared to subgiants are also not unusually good preservers of Li, implying that the boundary between Li-rich and not Li-rich may be lower, or even substantially lower, than A(Li) = 1.5 dex for giants in NGC 188, a pattern exhibited by other clusters of slightly younger age (Twarog et al.\citealt{Twarog20}, Anthony-Twarog et al.\citealt{Anthony20}, Anthony-Twarog et al.\citealt{Anthony21}).

To place the evolution of the progenitors of the NGC 188 giants in context, it is important to realize that nearly all low mass dwarfs deplete their surface Li abundance as they evolve during the MS, and that evidence from a variety of angles suggests that rotational mixing induced by angular momentum loss (Endal \& Sofia\citealt{Endal76}; Pinsonneault et al.\citealt{Pinsonneault89, Pinsonneault90}) is the dominant Li-depleting mechanism. We briefly summarize some of that evidence here; for more thorough discussion see C17 and Deliyannis et al.\citet{Deliyannis19}. Li, beryllium (Be, this symbol shall refer to the only stable isotope, $^9$Be, in contradistinction to $^7$Be, which shall always be labelled as $^7$Be), and boron (B) survive to progressively greater depths, so their ratios provide especially sensitive and robust probes of physical processes occurring in the outer layers of stars that may vary with depth. The correlated depletion of Li and Be in F dwarfs in the field (Deliyannis et al.\citealt{Deliyannis98}; Boesgaard et al.\citealt{Boesgaard01}) and in six open clusters\citet[Boesgaard et al.][]{Boesgaard02,Boesgaard04} is in excellent agreement with models of rotational mixing (Deliyannis \& Pinsonneault\citealt{Deliyannis93b,Deliyannis97}; Charbonnel et al.\citealt{Charbonnel94}), and argues strongly against other mechanisms proposed to explain the Li Dip. Similarly, the correlated depletion of Be and B in F dwarfs\citet[Boesgaard et al.][]{Boesgaard98,Boesgaard05b} also agrees nicely with similar models\citet[Boesgaard et al.][]{Boesgaard16}. Furthermore, F dwarfs in young open clusters ($\sim$ 100 Myr) show no correlation between A(Li) and {\it v} sin {\it i}, but a tight correlation forms as stars age, that becomes steeper with age, at least through the age of the Hyades ($\sim$ 650 Myr); this provides a direct correlation between Li depletion and spindown\citet[Steinhauer \& Deliyannis][]{Steinhauer22}. The early formation of the Li Dip also favors rotational mixing since other proposed mechanisms act later\citet[Steinhauer \& Deliyannis][]{Steinhauer04}.  But Li depletion due to rotational mixing is not confined to F dwarfs. Contrary to long-held beliefs, late A dwarfs have been discovered to spin down and deplete Li concurrently\citet[Deliyannis et al.][]{Deliyannis19}. Short-Period-Tidally-Locked-Binaries (SPTLBs), posited to be at least partly immune to rotational mixing during MS evolution, show higher A(Li) in normal, single late-F, G and K dwarfs, providing evidence of rotationally-induced Li depletion in these stars. This mechanism may be relevant to star 4705 (below), so we expound a bit further. The idea is that binaries observed today with orbital periods less than about 8 d were locked tidally and synchronized during the early pre-MS\citet[Zahn \& Bouchet][]{Zahn89} before the stellar interior was sufficiently hot to destroy Li. Therefore, they will not experience the spin-down, mixing, and Li depletion that single stars will experience later on, and thus SPTLBs could preserve their Li better than single stars do, though other possible complications may mean that some SPTLBs do deplete Li (D90, Deliyannis\citealt{Deliyannis90b}). SPTLBs with higher-than-normal Li are known in the Hyades, the turnoff of the 4-Gyr-old M67, and the field (Soderblom et al.\citealt{Soderblom90}; Thorburn et al.\citealt{Thorburn93}, Deliyannis et al.\citealt{Deliyannis94}, Ryan \& Deliyannis\citealt{Ryan95}). Other effects traced by Li may also be important for young, rapidly rotating G and K dwarfs that have higher A(Li) than stars rotating more slowly: there is evidence that such stars have inflated radii\citet[Jackson et al.][]{Jackson16, Jackson18, Jackson19}, caused by the rapid rotation and the action of magnetic fields (Feiden \& Chaboyer\citealt{Feiden14}; Somers \& Pinsonneault\citealt{Somers15}). So, uniformly, from late A dwarfs through K dwarfs, rotational mixing is the primary Li depletion mechanism. Subgiants in NGC 188 evolved out of turnoff stars that have A(Li) $\sim$ 2.4 dex, which is depleted nearly a factor of 10 if the initial abundance was similar to meteoritic ($\sim$ 3.3 dex), with rotational mixing the likely cause of the Li depletion. Then, the Li in the entire Li preservation region would also be depleted due to rotational mixing, as evidenced directly by the declining subgiant A(Li).

So what is ``Li-rich'' in NGC 188? Two of the four stars with detected Li (stars 4705 and 5027) already meet the canonical criterion of A(Li) $>$ 1.5 dex. The A(Li) = 1.17 dex of the yellow giant 4346 is far greater than our upper limits for RGB stars of similar luminosity and, in that sense, this star might be labelled ``Li-rich''. However, a great deal depends on how that star has reached this position in the CMD, as discussed below.  Figure \ref{fig:CMD_ALi} suggests that star 6353 is a RC star. (A low $^{12}$C/$^{13}$C ratio could support this suggestion but we are unaware of any C-ratio studies in giants of NGC 188.) Given that the subgiant A(Li) steadily declines  to levels below 1 dex as the stars evolve (Figure \ref{fig:subgiantLi}), that our Li upper limits continue to decline with lower $T_{\rm{eff}}$ up the RGB to a level as low as as A(Li) $<$ -1.2 dex, and that star 6353 not only has a higher A(Li) than these RGB progenitor analogs but also higher than four of the five other apparent RC stars, we propose that star 6353 is ``Li-rich'' even though its A(Li) is ``only'' 0.60 dex. If the concept of Li-richness identifies stars that have more Li than they could have preserved following MS evolution, then this star qualifies. Rather than simply preserving its MS Li, it is more likely that this star has been enriched in Li, either through internal or external means. 

By allowing for the possibility that some field stars with A(Li) $<$ 1.5 dex are Li-rich in this expanded sense, the fraction of Li-rich stars among the various red giant populations is likely to be higher than indicated in the Introduction. In addition to the ongoing cluster analyses by Twarog et al.\citet{Twarog20}, Anthony-Twarog et al.\citet{Anthony20}, and Anthony-Twarog et al.\citet{Anthony21}, support for a more malleable definition of Li-rich has emerged from a number of sources. Based on an extensive sample of GALAH giants, Kumar et al.\citet{Kumar20} find that A(Li) in the RC phase is higher than in the RGB stage and propose that Li production is ubiquitous in low-mass stars. This hypothesis has been discussed in several recent publications. Magrini et al.\citet{Magrini21} derive Li abundances and upper limits for RC and RGB stars which are open cluster members and find that 35\% of RC stars have similar or higher A(Li) than the RGB stars. They suggested possible Li production as the explanation but could not confirm that Li production is ubiquitous. By using a larger sample of 1848 field giants, Zhang et al.\citet{Zhang21} find that zero-age sequence, core-helium-burning low mass stars have higher A(Li) than stars above the RGB bump (Thomas Peak) by a value of $\sim$ 0.6 dex, and suggest that the helium flash may produce moderate Li enhancement while super Li-rich stars may require other mechanisms. Chaname et al.\citet{Chaname21} analyzed the GALAH Li field-star data, taking into account the distribution of progenitor masses, and stressed the need to account for population effects when comparing field RC and RGB stars. They find that standard models alone (no rotation, diffusion, magnetic fields, etc.) can account for the moderate A(Li) in RC stars, without the need for a Li production mechanism. They predict that higher mass RC stars should have higher A(Li), and cite some supporting evidence. We stress that, in addition to consideration of the mass-dependence of the progenitor, the MS evolution of the Li abundance prior to entering the subgiant branch must be considered, as we argue in Twarog et al.\citet{Twarog20} and again in this work. For reasons already noted, clusters, rather than field stars, are invaluable in this endeavour.

\subsection{The Origin of Lithium In Our Li-Rich Stars}

Some properties of the RG sample in NGC 188 are striking, in particular that 4 of 29 members (14\%) are Li-rich. This is a much higher rate than in the field ($<$ 1\%) (Casey et al.\citealt{Casey19}; Deepak \& Reddy\citealt{Deepak19}), even if we only count the two stars (7\%) with A(Li) $>$ 1.5 dex. Assuming that our higher rate isn't merely an artifact of small number statistics, why would Li-rich stars occur at a higher rate in NGC 188?

It is of interest to compare to the rate of Li-rich stars in other open clusters. In the current discussion, ``Li-rich'' means A(Li) $>$ 1.5 dex, but ``Li-rich-in-context'' means A(Li) is less than 1.5 dex but abnormally high A(Li) compared to other relevant cluster stars, as argued here for two of the giants in NGC 188. To minimize the possibility of confusing subgiants still undergoing Li dilution from truly Li-rich giants, we also define ``giants'' to be those stars on the proper RGB, i.e. brighter than the reddest portion of the subgiant branch.  For example, this definition eliminates several subgiants in NGC 2506 that are clearly still undergoing dilution \citet[Anthony-Twarog et al.][]{Anthony18}, star 87 of IC 2714 of Delgado Mena et al.\citet{DelgadoMena16} that has A(Li) = 1.54 dex but a temperature of 5377 K on the subgiant branch, and so on. Table \ref{tab:Li_OC} summarizes a number of previous studies. The number of giants indicated corresponds to the number observed and reported in each study that meet the above criterion, except for Andrievsky et al.\citet{Andrievsky99} who report on only one star, a Li-rich giant. We estimate that the cluster may have about 10 giants, so at least 10\% are Li-rich. Delgado Mena et al.\citet{DelgadoMena16} observed 67 subgiants/giants in 12 open clusters, but only 32 giants as we defined them above, and found no Li-rich giants but 3 Li-rich-in-context (0\% and 9\%, respectively). The remaining studies in Table \ref{tab:Li_OC} attempted to observe the vast majority of candidate giant members in each cluster. In total, these studies found 2 out of 139 giants to be Li-rich (1.4\%), a rate that is somewhat closer to those in the field than that for NGC 188, and also 1 star that was Li-rich-in-context. Thus, 3 out of 139 giants (2.2\%) are either Li-rich or Li-rich-in-context. In combining these statistics with the other two studies in Table \ref{tab:Li_OC}, we note that Delgado Mena et al.\citet{DelgadoMena16} observed 3 giants in NGC 3680, of which two are at the Thomas Peak and 1 is just a bit below. They also observed a few stars in IC 4756 that are too hot to be labeled ``giants'' by the above criterion. It is possible, perhaps even likely, that these 3 stars were observed by Anthony-Twarog et al.\citet{Anthony09}, but we cannot be sure. The two studies find similar A(Li) for stars at that specific phase of evolution which are neither Li-rich nor Li-rich-in-context. We include the 3 stars of Delgado Mena et al.\citet{DelgadoMena16} as distinct from the 9 in Anthony-Twarog et al.\citet{Anthony09}, which introduces a risk of slightly underestimating the fraction of Li-rich or Li-rich-in-context stars. Thus, adding together all the stars of Table \ref{tab:Li_OC}, there are 3 Li-rich stars out of 181 (1.7\%), and 7 stars out of 181 (3.9\%) that are either Li-rich or Li-rich-in-context. Adding the present statistics from NGC 188, the totals become 5 Li-rich giants out of 210 (2.4\%), and 11 stars out of 210 (5.2\%) that are either Li-rich or Li-rich-in-context. By comparison, Deepak \& Reddy\citet[][from the GALAH survey]{Deepak19} found only 0.64\% of Pop I red giants below $2M_{\odot}$ to be Li-rich while Gao et al.\citet{Gao19} found 1.29\% of all giants to be Li-rich. For completeness, Kirby et al.\citet{Kirby16} found only 0.3$\pm$0.1\% of globular cluster giants to be Li-rich.

\begin{table*}
	
\caption{Li-rich giants in Open Clusters}
\label{tab:Li_OC}
\begin{tabular}{ccccc}
\hline

Cluster & number & number & number & reference \\
& of & Li-rich & Li-rich & \\
& giants & & in context & \\

\hline
IC 4756 & $\sim$ 10 & 1 & 0 & Andrievsky et al.\citet{Andrievsky99} \\ 
NGC 3680 & 9 & 0 & 0 & Anthony-Twarog et al.\citet{Anthony09} \\
NGC 6253 & 17 & 0 & 0 & Cummings et al.\citet{Cummings12} \\
NGC 6819 & 40 & 1 & 0 & Anthony-Twarog et al.\citet{Anthony13}\\ 
NGC 2506 & 48 & 1 & 0 & Anthony-Twarog\citet{Anthony18} \\ 
NGC 2243 & 25 & 0 & 1 & Anthony-Twarog et al.\citet{Anthony20,Anthony21} \\ 
12 OCs & 32 & 0 & 3 & Delgado Mena et al.\citet{DelgadoMena16} \\
\hline

\end{tabular}
\end{table*}

Returning to a point of emphasis in the current analysis, note that the Li-rich giants and giants with measurable Li in other open clusters discussed to date (for example, the ones in Table \ref{tab:Li_OC} or in Magrini et al.\citealt{Magrini21}) all come from MS turnoff stars above (more massive than) the Li Dip. By contrast, turnoff stars from NGC 188 come from MS stars below (less massive than) the Li Dip, so the mechanism(s) controlling Li evolution before and on the RGB may be distinctly different in NGC 188, i.e. the statistics and the Li-generating mechanism(s) may change dramatically with lower mass. Compounding the complexity, the giants in M67 and the older super-metal-rich clusters NGC 6253\citet[Cummings et al.][]{Cummings12} and NGC 6791 \citet[Boesgaard et al.][]{Boesgaard15} have no detectable Li, but come from MS stars that populate the Li Dip, placing them at a severe Li disadvantage well before leaving the main sequence and substantially lowering the boundary for Li-rich-in-context.

As noted earlier, Figure \ref{fig:CMD_ALi} shows that the majority of NGC 188 stars define a precise RGB. Some stars lie slightly off of this fiducial (orange circles). This is not due to photometric error; the same stars are slightly off in the same way based upon either Gaia photometry or our own extended Stromgren photometry \citet[Twarog et al.][]{Twarog21}. As expected, the RC is slightly to the blue of the RGB for RGB stars of similar luminosity (blue circle). {\it Strikingly, none of the four Li-rich stars lies on the fiducial sequence. One is classified by CMD poisiton as a RC giant, just slightly fainter and redder than the other presumed RC stars. {\bf All three} of the others are well to the blue of the RGB.} We shall refer to these three as ``yellow giants'' (YG). These unusual CMD positions may be telling us something about the origin of the Li in these stars. Note that while the vast majority of Li-rich field stars are RC stars, only 1 of our 4 Li-rich stars in NGC 188 is a probable RC star. We use the qualifier ``probable'' to stress again that position in the CMD does not necessarily guarantee knowledge of evolutionary state, as exemplified by star WOCS 7017 in NGC 6819, a RC star fainter than the normal RC and redder than the RGB. In fact, we cannot altogether rule out that this star might be on the RGB; observations of the $^{12}$C/$^{13}$C could help distinguish between RGB (high ratio) and RC (low ratio). We consider each of the four stars and their possible origins in turn.

\subsubsection{RC Star 6353}

Possible sources of Li for star 6353 as a single star include the $^7$Be transport mechanism occurring either at the He-core flash, with thermohaline mixing near the RGB tip, or at the Thomas Peak, and engulfment. Using Li depletion timescales, Casey et al.\citet{Casey19} argue that neither engulfment at a random time on the RGB, nor Li production near the RGB tip or at the Thomas Peak can be the dominant mechanism that explains the Li-rich RC stars. They acknowledge our lack of understanding of the complexities involved in a (likely turbulent) He-core flash, to which we add core rotation and asymmetry. Nonetheless, they propose that tidal spin-up from a binary companion \citet[Fekel \& Balachandran][]{Fekel93} can explain most of the Li-rich stars. While that may well be true, since star 6353 is only one star, any of the single-star mechanisms listed above might apply. One inconvenient and confounding finding is that the carbon abundances and $^{12}$C/$^{13}$C are very similar in Li-rich (with A(Li) up to 3.1 dex) and Li-poor K giants, and consistent with first dredge-up in both cases. It would be informative to learn how the C-ratio in this star compares to other RC and RGB members of the cluster.  

Among many other possibilities\citet[Sun][]{Sun21} is the option that star 6353 began life as a binary. One star evolved to create a HeWD and the other star merged with the HeWD as it evolved up the RGB, ending up on the RC.  Following the merger, a hot helium shell surrounds the hybrid/merged core and the SCZ deepens briefly and results in surface $^7$Li enhancement via the $^7$Be-transport mechanism. For this to work requires appropriate masses for both stars, and significant mass loss given star 6353's location in the lower right portion of the RC, which could happen through ejection of the common envelope. 

A number of observational tests exist that might distinguish between engulfment and the mixing scenarios. Like Li, Be and B also experience depletion due to rotational mixing, subgiant dilution, and the $^7$Be-transport mixing mechanism that creates Li. But engulfment brings to the star fresh unburnt Be and B. So stars that enriched Li through engulfment should have higher Be and B than those that did so via the $^7$Be-transport mechanism. Be and B observations in giants are difficult, but perhaps not impossible.

\subsubsection{YG Star 4705}

Star 4705 is the binary yellow giant. It might be tempting to assume that the system is composed of a red giant and white dwarf which earlier had transferred mass, Li, and angular momentum to its now more luminous companion, especially given the absence of a second set of absorption lines due to the companion to the red giant among our Hydra spectra (See, e.g. Figure \ref{fig:comb_ALi}). However, this scenario is excluded since Kaluzny\citet{Kaluzny90} discovered the star to be a variable (V11). Based upon its CMD location blueward of the red giant branch and a $V$ mag decline of almost 0.4 mag on one CCD frame, Kaluzny\citet{Kaluzny90} presciently proposed that the star was an eclipsing binary composed of a normal red giant and a main sequence star near the top of the turnoff. In later analysis of an expanded but still incomplete light curve, Mazur \& Kaluzny\citet{Mazur90} demonstrated that star 4705 showed definite variability at the level of 0.06 mag over a timescale of just under two weeks, leading to classification as a possible RS CVn-type variable, in addition to its binary nature. This classification was enhanced by the detection of star 4705 as an X-ray source by Gondoin\citet{Gondoin05}, implying magnetic activity triggered by the combination of rotation with convection. Zhang et al.\citet{Zhang02} found small amplitude variability in star 4705 but proposed a period of 1.2433 days, now known to be an order of magnitude too small. The binary nature of the system was confirmed by Geller et al.\citet[][= G09]{Geller09}, who also found that the 0.4 mag drop in brightness detected by Kaluzny\citet{Kaluzny90} coincided in time with the eclipse phase (t $\sim$ 0.02) implied by their radial-velocity curve. With an orbital period and eccentricity of 35.178 days and 0.487, respectively, the masses of the system members were calculated as 1.14 $M_{\odot}$ and 1.09 $M_{\odot}$, placing the two stars on the giant branch and the turnoff, respectively. A similar deconvolution is discussed in  Vats et al.\citet{Vats18} where improved X-ray observations via {\it Chandra} again lead to classification of the system as an active binary. Unfortunately, part of the discussion placing star 4705 in the larger context of X-ray sources is hampered by the inclusion of the flawed rotation period derived by Zhang et al.\citet{Zhang02}.

\begin{figure}

	\centering
	\includegraphics[width=0.35\textwidth, angle=270]{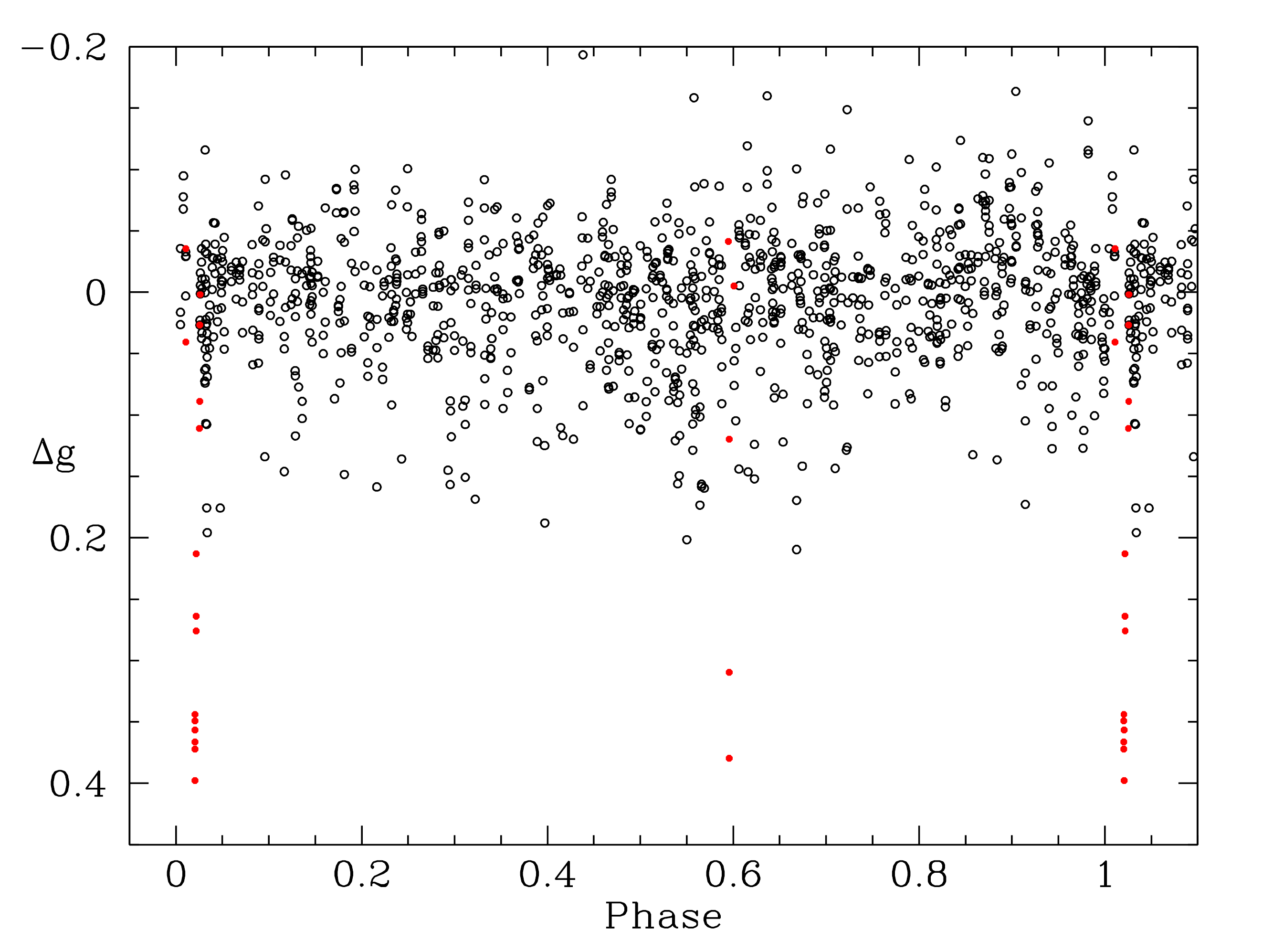}
	\caption{Light curve of Star 4705, with apparent eclipses indicated in red.}
	\label{fig:lightcurve}

\end{figure}

To resolve some of the issues tied to the binary/rotational nature of the system, we turn to the variable star database compiled by the All-Sky Automated Survey for SuperNovae (ASAS-SN Shappee et al.\citealt{Shappee14}) variable stars database (\url{https://asas-sn.osu.edu/variables}). Star 4705 is the cataloged star, ASASSN-V J004522.63+851238.2/NSV 15158, classed as ROT, meaning the variability is likely due to rotationally-induced brightness modulation, with an amplitude of 0.17 mag and a period of 17.66357 days\citet[Jaysinghe et al.][]{Jaysinghe09}. The rotation period is very close to $\frac{1}{2}$ of the orbital period and obviously consistent with the original suggestion of Mazur \& Kaluzny\citet{Mazur90}. While the system is clearly a radial-velocity binary, the only indication to date of a possible eclipse comes from the single data point noted by Kaluzny\citet{Kaluzny90}. With the extensive data available from ASAS-SN, we can attempt to remedy this deficiency. {\it g} mag photometry for star 4705 collected over a 1000-day span was collated and refined by the removal of all observations with errors of 0.10 mag or larger. The variation in $g$ due to rotation was then removed from the data using a periodicity of 17.6636 days and the corrected magnitudes phased using the periodicity and timing based upon the orbital information derived by G09. The resulting light curve is shown in Figure \ref{fig:lightcurve}, where the data coupled to the predicted eclipse times are plotted in red. While one might question the reality of the eclipse at phase t = 0.5957 due to the limited sequence of points obtained within only a single orbital cycle, the same argument cannot be made of the eclipse at phase t = 0.0207, composed of photometry from three distinct orbital cycles, in excellent agreement with the original claim of Kaluzny\citet{Kaluzny90}. Based upon the limited time span of the data points, the total eclipse times fall below 13 hours and 5 hours for the phase 0.02 and 0.60 eclipses, respectively.

A possible solution to our lack of spectroscopic evidence for the companion lies in the derived binary orbits of NGC 188 (G09). Given the placement (by G09, described above) the stars on the giant branch and the turnoff, the luminosity of the secondary is hardly negligible.  G09 observed the star as an SB2 over multiple orbits at multiple locations within the orbit, producing excellent orbital parameters. By coincidence, we acquired our Li data at orbital phase 0.52 $\pm$ 0.03, where nearly the entire error is propagated from the uncertainty in the period. This is close to phase 0.59 when the stars eclipse\citet[Kaluzny][]{Kaluzny90} and thus have the same radial velocity, but sufficiently far from that specific point that it is likely our spectra received the full (uneclipsed) light from both stars. It would thus seem that our spectrum is a composite of the spectra from both stars, with no detected velocity shift. This also explains the oddity that the $\sigma_{\mu}$ in the [Fe/H] of this star is the second highest in our sample. Excepting the first and third highest, all other stars have substantially lower $\sigma_{\mu}$ and 52\% of the stars have a $\sigma_{\mu}$ that is less than half of that of star 4705. Since we are deriving [Fe/H] from multiple lines that may depend on $T_{\rm{eff}}$ in different ways, a high $\sigma_{\mu}$ is not surprising from a composite spectrum being treated as a spectrum of a single star with incorrect parameters.  Related to this, note also that the syntheses of the non-Li lines for the three other stars of Figure \ref{fig:comb_ALi} fit the data better than for star 4705. (The five most prominent non-Li lines are Fe I lines.)

What about the Li?  We have explored various possibilities, assuming one star is near the well-populated turnoff and one is near the lower part of the RGB. In each case, we have assumed combinations of $V$ and $B-V$ for the individual stars that reproduce the observed combined $V$ and $B-V$. Since this star is an X-ray variable and possibly an RS CVn system (Gondoin\citealt{Gondoin05}, Zhang et al.\citealt{Zhang02}), and since such stars are sometimes known to be red stragglers\citet[Geller et al.][]{Geller17}, we consider also the possibility that the giant is just slightly to the red of the RGB. At least one such star is known in NGC 188 (star 1141 in our sample) and another red straggler is known but without detected X-ray emission (star 3118).  Both stars are near the proposed location in the CMD for the red giant and, in fact, their $V$ magnitudes straddle our proposed $V$ for this star.

For each combination, we synthesized a spectrum for each star and then combined the syntheses using luminosity ratios inferred from the stellar R magnitudes, which represent our spectral region well. The R magnitudes were taken from stars occupying the assumed CMD positions of the proposed stars. The renormalized combined synthetic spectrum could then be compared to the observed one. A variety of combinations was considered, and Table \ref{tab:synth_4705} shows a subset of illustrative examples. Star 1 is the turnoff star and star 2 is the red giant.

\begin{table*}

\caption{Possible combinations for Star 4705}
\label{tab:synth_4705}
	\begin{tabular}{cccccccc}
	\hline
	{Case} & $V_1$ & $(B-V)_1$ & $V_2$ & $(B-V)_2$ & $A(Li)_1$ & $A(Li)_2$ & Comment \\
	\hline
	A & 14.85 & 0.74 & 14.52 & 1.17 & 2.5 & 0.95 & top of turnoff, red straggler giant \\
	B & 14.85 & 0.74 & 14.52 & 1.17 & 2.6 & 0.8 & top of turnoff, red straggler giant \\
	C & 15.30 & 0.70 & 14.27 & 1.09 & 2.5 & 1.25 & fainter turnoff, fiducial RGB \\
	D & 15.30 & 0.70 & 14.27 & 1.09 & 3.1 & 0.8 & fainter turnoff, fiducial RGB \\
	\hline

\end{tabular}

\end{table*}

Cases A and B employ a star at the brightest portion of the turnoff; then, the combined magnitude/color restriction forces the red giant to be a red straggler. Cases C and D consider a fainter turnoff star; in this case, the red giant falls right on the RGB fiducial. For each turnoff position, we assumed a Li abundance consistent with those observed in other turnoff stars, namely A(Li) $\sim 2.5$ dex, and varied the giant's A(Li) to match the observed spectrum (cases A and C). The question then becomes, is the giant's A(Li) reasonable?  We also reversed the question and gave the giant an A(Li) = 0.8 dex, consistent with the upper limits seen in the lower RGB, and varied the turnoff star's A(Li) to match the observed spectrum (cases B and D). The question now becomes, is the turnoff star's A(Li) reasonable?

\begin{figure}

	\centering
	\includegraphics[width=0.45\textwidth]{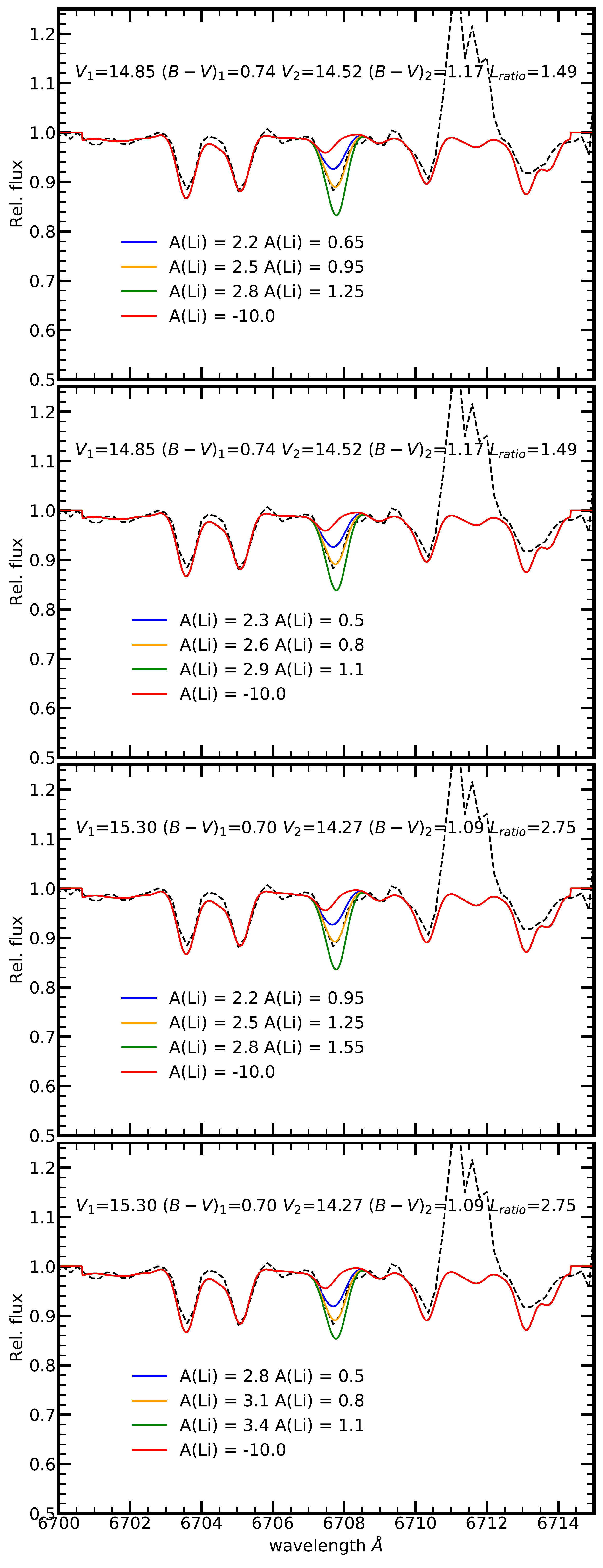}
	\caption{Syntheses for star 4705 under the four cases in Table \ref{tab:synth_4705}. The same symbols are used as in Figure \ref{fig:comb_ALi}.}
	\label{fig:star4705}

\end{figure}

Figure \ref{fig:star4705} shows the syntheses for these four cases compared to the observed spectrum. Note that the non-Li lines now fit much better than in Figure \ref{fig:comb_ALi}, arguably as well as for the other three stars, which supports the idea that one star is a turnoff star and the other is a red giant. In cases A and C, even though the turnoff star is assumed to have A(Li) = 2.5 dex, a fainter star (case C) contributes less flux to the combined spectrum so the giant must have higher A(Li) than in case A to match the observed Li line.  In both cases A and C, the A(Li) of the red giant appears to be marginally too high compared to the giants at the base of the RGB; however, this is consistent with the argument that RS CVn stars have higher-than-normal A(Li)\citet[Randich et al.][]{Randich94}. Another possibility is that this binary has some of the Li-preserving capability of SPTLBs (more on this below).

In cases B and D, the turnoff star has a higher A(Li) than is seen in other turnoff stars. Could the turnoff star be a better Li preserver than the other turnoff stars?  As mentioned above, some SPTLBs are known to be better preservers of Li than single stars, and the 4-Gyr-old M67, in particular, has a SPTLB with higher-than-normal A(Li) in both components of the binary\citet[Deliyannis et al.][]{Deliyannis94}. At over 35 d, the orbital period of star 4705 is too high to classify it as a proper SPTLB. However, the ratio of the orbital and rotational periods is 2:1 to at least two significant digits, suggesting some tidal interaction. The eccentricity is high (0.487 $\pm$ 0.005, G09), so there are times when the two stars are closer than they would be in a circular orbit with the same period, allowing for greater tidal interaction. If the 2:1 ratio was established during the early pre-MS when both stars were large, as opposed to recently, then, like SPTLBs, one or both stars could have avoided some or all of the spin-down and Li depletion that single stars experience. It is thus at least plausible that one or both components of star 4705 have preserved more Li than single cluster stars at similar evolutionary states.

Note that all four cases rely on the presence of Li in the red giant spectrum.  If instead the giant is devoid of Li, the turnoff star would require an A(Li) higher than meteoritic, which seems unlikely.  

We thus favor plausible scenarios, such as the ones described above, where one or both stars have A(Li) near or a bit above the A(Li) of single cluster stars in similar evolutionary states.  Further progress might be possible by the acquisition of spectra obtained at a point in the orbit when both stars are individually observable. 

\subsubsection{YG Stars 4346 and 5027}

\begin{figure}
	\centering
	\includegraphics[width=0.45\textwidth]{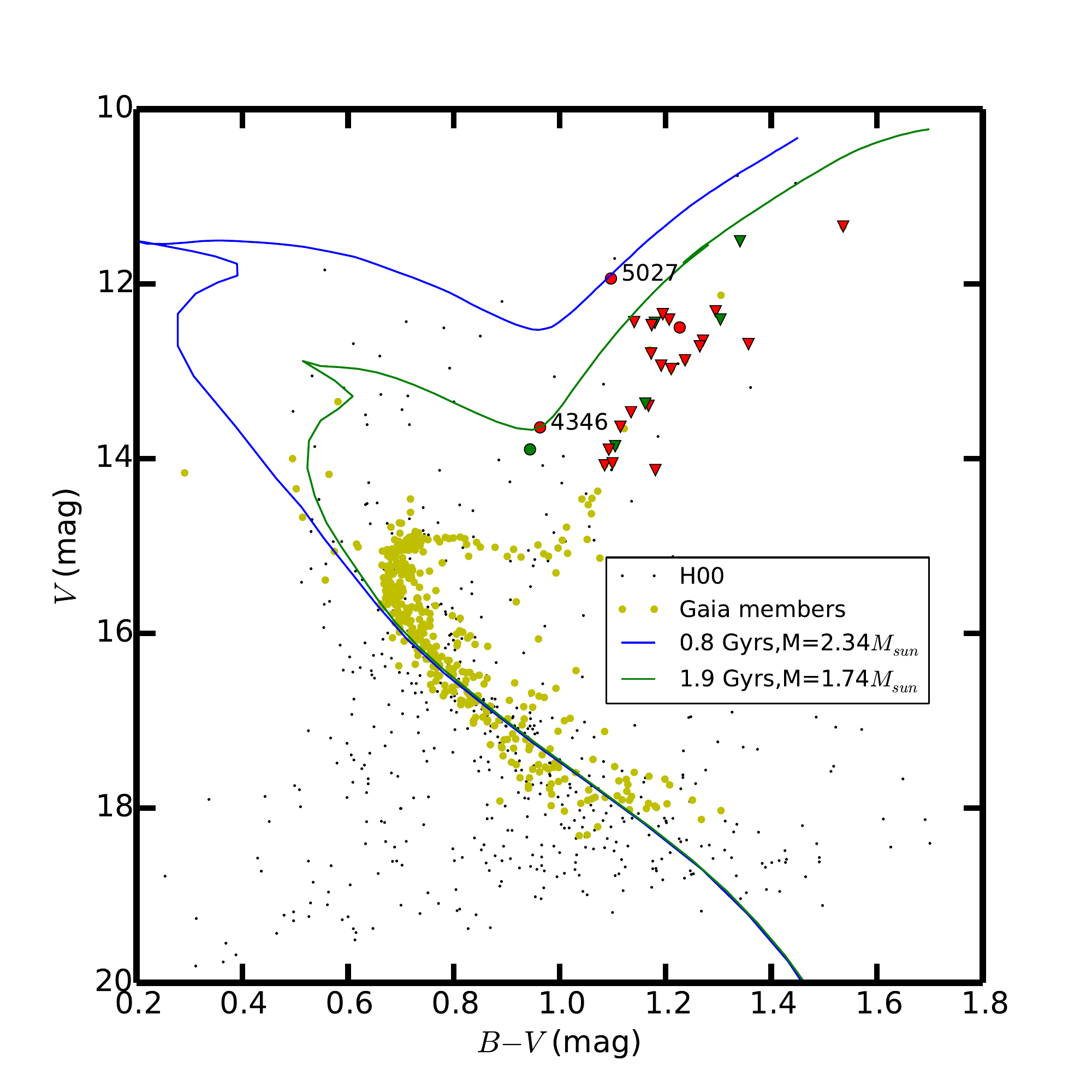}
	\caption{Yale-Yonsei isochrones with [Fe/H] = +0.06 for stars 4346 and 5027.}
	\label{fig:iso_4346_5027}
\end{figure}

Positioned blueward of the RGB like star 4705, these stars are classified as yellow giants. Critically, however, stars 4346 and 5027 are spectroscopically tagged as single-star cluster members. Taken at face value and ignoring for now the option of anomalous evolution, we can predict the masses of the individual stars. Figure \ref{fig:iso_4346_5027} shows the appropriate isochrones that can overlay each star. 

Star 5027, the brighter YG, is optimally matched by an isochrone of age 0.8 Gyrs, implying a mass of 2.34 $M_{\odot}$, while star 4346 fits the analogous parameters of 1.9 Gyr and 1.74 $M_{\odot}$. Given a total absence of evidence for multiple generations of stars within NGC 188, the most plausible option is the formation of a higher mass star via the merger of a close main-sequence binary pair, a common scenario linked to the presence of blue stragglers in open clusters. If both stars are evolved, merged BS, they both find themselves in the rapidly evolving giant stage\citet[Sills et al.][]{Sills09} when only 21 stars are known to be BS members of NGC 188\citet[Gosnell et al.][]{Gosnell15}. Furthermore, at least 17 (81\%) of the 21 known BS in NGC 188 are binaries, yet both yellow giants are single. Of the 17 binaries, 15 are SB1 with 14 thought to have been formed by mass transfer, and the other two are SB2 with unknown formation history. At least 7 of the mass-transfer BS are known to include WD\citet[Gosnell et al.][]{Gosnell15, Gosnell19}, both He and C/O WD, implying some of the WD passed only through part of the RGB phase before mass transfer while others made it to the AGB phase. Again, the single-star nature of these YG implies a merger scenario.

Although the evolution of a BS originating from a merger will be slightly different than that of a single star of similar mass, the evolutionary similarities are sufficient to allow some possible conclusions. In particular, the inferred masses are either less than (star 4346) or close to (star 5027) twice the current turnoff mass to within the errors, consistent with the idea of a recent merger as one star begins to leave the MS and expands. Adopting the current mass of the turnoff stars to contrain one of the original members of the pair, one finds for star 4346 and a composite mass of 1.74 $M_{\odot}$ that the individual masses are 1.1 $M_{\odot}$ for the turnoff star and 0.64 $M_{\odot}$ for the other star. For star 5027, both pre-merger stars must have been turnoff stars with masses just above 1.1 $M_{\odot}$.

How does this scenario affect the observed Li abundance? Keeping in mind the multiple evolutionary pathways open to such a system given an array of possible orbital and rotational parameters, in the simplest case for star 4346, the fainter star with A(Li) = 1.17 dex, the low mass secondary will have destroyed all of its Li quickly through standard pre-MS destruction alone, which must occur even in a SPTLB, so all the Li observed in the merged system must come from the turnoff star. If the merger process is sufficiently quiet so as not to destroy any Li, then there are two dilution factors that must be taken into account: a) the fact that a 1.74 $M_{\odot}$ merged star has the reservoir of only a 1.1 $M_{\odot}$ MS star (dilution by $\sim$ 0.2 dex), and b) subgiant dilution (dilution by $\sim$ 1.5 dex by the current stage of star 4346, Charbonnel \& Lagarde\citealt{Charbonnel10}).  This is a total dilution of 1.7 dex, implying a pre-merger A(Li) for the turnoff star of 2.85 dex. For a SPTLB, such an A(Li) is possible! It is slightly higher than the A(Li) $\sim$ 2.5 dex seen in current turnoff stars, and lower than meteoritic. 

Given the caveats noted above, for the brighter YG, star 5027 with A(Li) = 2.04 dex, both pre-merger stars must be turnoff stars and, for optimistic Li preservation assumptions, both are allowed to have their maximally preserved A(Li). This means a dilution factor on the MS of 0.0 dex but subgiant dilution near maximum, or 1.8 dex, for a total dilution of 1.8 dex. This implies a pre-merger A(Li) for each turnoff star of $\sim$ 3.8 dex, prohibitively high given it is 0.4-0.5 dex above the presumed initial cluster value, unless the cluster formed with an unexpectedly and correspondingly high A(Li). However, young clusters, whose initial A(Li) may be evaluated more easily since some of their stars are minimally depleted in Li, show no evidence of such large and super-meteoritic variations in initial Li (C17). Thus, the simplest alternative for the high A(Li) remains the merger of two stars which individually have retained a higher than average Li abundance after passing through the subgiant branch and merging not long after this phase. The obvious candidate for such a scenario is a SPTLB composed of two stars of very similar mass. The evidence for anomalously high A(Li) for stars in such systems is discussed in detail in Deliyannis et al.\citet{Deliyannis19} and will not be repeated. Presumably, the merger process can occur in a variety of ways\citet[e.g. Nelson \& Eggleton][]{Nelson01}, including short or long common envelope phases, mixing that dilutes or destroys Li, and various mass loss and angular momentum transport scenarios. Without significantly more detailed constraints on these two stars, further speculation seems unwarranted (see also Sun\citealt{Sun21}).

It should be noted for completeness that star 5027 (but not star 4346) can also be fit by the RC portion of a Padova isochrone\citet[Bressan et al.][]{Bressan12} of $\sim$ 1 Gyr. If star 5027 is a RC star, this opens up the same Li-enchancement possibilities discussed above for star 6353. In addition, perhaps the Zhang et al.\citet{Zhang20} HeWD+RGB merger mechanism can produce a yellow giant sufficiently to the red of the current RC, keeping in mind the seemingly stringent constraint that if star 5027 is a RC star, it appears to have twice the turnoff mass. While this may be a possible explanation for star 5027, it is not so for star 4346. 

Future observations could help clarify the origin of the Li in these stars. For example, if dilution alone has changed the surface A(Li), then there would be corresponding dilution of Be and B and a decrease in the C-ratio. However, if the Li was first mixed and destroyed, destroying also a corresponding amount of Be and B larger than dilution alone, and then the $^7$Be-transport mechanism has created the observed Li (but not Be and B), then Be and B would be lower. Asteroseismology could determine masses and distinguish between RGB and RC internal structures. For example, Leiner et al.\citet{Leiner16} found a yellow straggler in M67 and used asteroseismic analysis to demonstrate that it is a helium burning giant with a mass of approximately twice that found at the cluster turnoff.

\section{Summary}       \label{sec:summary}

We present WIYN/Hydra spectra of 34 red giant stars that are proper-motion members or member candidates from P03 in the old open cluster NGC 188. Observations were made on two nights in one configuration. After considering our and G08's $V_{RAD}$ and multiplicity results, and combining with $\pi$, $\mu_{\alpha}$, $\mu_{\delta}$ information from Gaia DR2, we find 23 single members, 6 binary members, 1 binary nonmember, and 4 single nonmembers, and a mean cluster $V_{RAD}$ = -42.89 $\pm$ 0.16 km s$^{-1}$ ($\sigma_{\mu}$, $\sigma$ = 0.93 km s$^{-1}$).

We adopted the average of H00 and S99 photometry ($BVRI$, $U$ magnitude from S99), or the P03 photometry for stars not in S99 or H00, and used all ten possible color combinations from $UBVRI$ to derive an ``average'' $B-V$ to reduce the scatter in the $T_{\rm{eff}}$ derived from $B-V$. The log $g$ values were initially derived from a $Y^2$ isochrone with [Fe/H] = 0.00 dex and E($B-V$) = 0.09 mag and $\xi$ were calculated using the relationship from Carretta et al.\citet{Carretta04} for giants. We selected seven non-blended Fe I lines using the Arcturus spectrum, measured equivalent widths for each line for each star, and adopted the Kurucz\citet{Kurucz92} stellar atmosphere models to derive iron abundances using the {\it abfind} task in MOOG for each line. After excluding outlying lines and lines that are $T_{\rm{eff}}$, log $g$, or $\xi$ dependent, we used the surviving lines to calculate [Fe/H] strictly relative to the Sun (using solar gf-values) for each line for each star. We then derived linear weighted [Fe/H] in linear space for each star and for the whole cluster, $[Fe/H]_{188}$ = +0.084 $\pm$ 0.016 dex ($\sigma_{\mu}$, and $\sigma$ = 0.162 dex). We then iterated to refine these values, which led to our final cluster average [Fe/H] = +0.064 $\pm$ 0.018 dex ($\sigma_{\mu}$, and $\sigma$ = 0.177 dex).

The cluster's well-defined parameters and cool giants with the lowest gravities provided an opportunity to refine the line list near the Li 6707.8 \AA, which clearly showed too much absorption in these stars. We adjusted the gf-values of a few lines to match the stellar spectra at various $T_{\rm{eff}}$, under the conservative assumption that these stars have no Li; the presence of Li would imply that these gf-values should be even lower. This revised line list will enable more confident detection of Li in G- and K- dwarfs with extremely weak Li lines, and such detections may help elucidate the physical mechanisms responsible for Li depletion in low-mass stars.

Application of MOOG to the member giants using the revised line list led to Li detections in four giants, namely stars 5027 (A(Li) = 2.04 dex), 6353 (A(Li) = 0.60 dex), 4346 (A(Li) = 1.17 dex), and 4705 (A(Li) = 1.65 dex), with 3$\sigma$ upper limits for the rest. Canonically, field stars with A(Li) $>$ 1.5 dex are called Li-rich because of the expectation of $\sim$ 1.8 dex subgiant Li dilution from a presumed initial A(Li) = 3.3 dex.  However, rotational mixing depletes the surface Li of dwarf stars during the MS and possibly beyond. In the case of NGC 188, since the stellar A(Li) at the turnoff are $\sim$ 2.5 dex and then vanish to much lower levels than 1.5 dex as subgiants evolve to the base of the RGB, all four stars are Li-rich in this cluster's context. An incidence of 4 (or even just 2) Li-rich stars in a sample of 29 stars is far higher than what recent large surveys have found in the field. All four stars lie either slightly or substantially away from the cluster fiducial sequence, possibly providing clues about their Li-richness.

Our only Li-rich star that is a binary, star 4705, is composed of a turnoff star and a star at the base of the RGB.  By misfortune we observed this SB2 near eclipse so the unobstructed spectra of both stars appear combined in our data. Composite synthesis suggests a possible simple solution where the RGB component has a low A(Li) that is approximately consistent with the upper limits seen in single RGB stars at the same evolutionary stage, and the turnoff star has slightly higher A(Li) than is seen in single turnoff stars, where tidal interactions might have helped preserved more Li in this star during the MS.  

Star 6353 appears to be a RC star, although it is slightly redder and fainter than the other RC stars. This could indicate a more severe mass loss at or near the tip of the RGB, as was clearly the case for star WOCS 7017 of NGC 6819, and is consistent with the appearance of surface Li from the $^7$Be transport mechanism following a He-core flash. We cannot rule out other possibilities such as planet engulfment, and discuss how future studies of the star's CNO abundances, $^{12}C/^{13}C$ isotope ratio, and Be and B abundances may help distinguish between possible scenarios.

Stars 4346 and 5027 are yellow giants. Since only 21 stars are known to be blue stragglers of NGC 188, evolutionary time scales and other arguments suggest it is unlikely that these two stars are evolved blue stragglers. Besides, BS in other clusters are Li-poor, and no mechanism has been proposed where BS can create Li, although we speculate about a few such possibilities. A more promising scenario might be the recent merger of a contact binary, where at least one star originates at the well-populated turnoff. Under optimistic but not necessarily unrealistic Li preservation assumptions, both during the MS and the merger event, it might be possible to explain star 4346, but not star 5027.  On the other hand, star 5027 (but not star 4346) might be a RC star, in which case the possible scenarios for star 6353 may apply here also. We again propose potentially helpful observations.

\section*{Acknowledgements}

We thank an anonymous referee for their constructive comments which helped to improve this work. We thank Alison Sills and Emily Leiner for very insightful discussions about blue and red stragglers and stellar mergers. NSF support for this project was provided to C.P.D. through grant AST-1909456. Q.S. thanks support from the Shuimu Tsinghua Scholar Program. We would also like to thank the WIYN 3.5m staff for helping us obtain excellent spectra, and the European Space Agency (ESA) for Gaia data (\url{https://www.cosmos.esa.int/gaia}).

\section*{Data Availability}

The spectra used in this study are proprietary. All other data are included or obtained from published studies as referenced in the article.



\bibliographystyle{mnras}


\bsp	
\label{lastpage}
\end{document}